\documentclass[journal]{IEEEtran}
\IEEEoverridecommandlockouts
\usepackage{cite}
\usepackage{amsmath,amssymb,amsfonts}
\usepackage[algo2e]{algorithm2e}
\usepackage{graphicx}
\usepackage{textcomp}
\usepackage{xcolor}
\usepackage{amsmath}
\usepackage{caption}
\usepackage{subcaption}
\usepackage{enumitem}

\SetKwInput{KwInput}{Input}                
\SetKwInput{KwOutput}{Output} 
\usepackage[utf8]{inputenc}
\usepackage[english]{babel}
\newtheorem{theorem}{Theorem}
\usepackage{algorithm,algorithmic,amsmath}
\usepackage{mathtools}
\usepackage{amsmath}
\usepackage{comment}
\def\BibTeX{{\rm B\kern-.05em{\sc i\kern-.025em b}\kern-.08em
    T\kern-.1667em\lower.7ex\hbox{E}\kern-.125emX}}
\usepackage{optidef}

\begin{document}

\title{Recursive Gaussian Process over graphs for Integrating Multi-timescale Measurements in Low-Observable Distribution Systems\\

}

\author{Shweta~Dahale,~\IEEEmembership{Graduate Student Member,~IEEE} and
       Balasubramaniam~Natarajan,~\IEEEmembership{Senior~Member,~IEEE}
\thanks{S. Dahale and B. Natarajan are with Electrical and Computer Engineering, Kansas State University, Manhattan, KS-66506, USA, (e-mail:
sddahale@ksu.edu, bala@ksu.edu). This material is based upon work supported by the Department  of  Energy,  Office  of  Energy  Efficiency  and  Renewable Energy  (EERE),  Solar  Energy  Technologies  Office,  under Award Number DE-EE0008767}}

\maketitle

\begin{abstract}
The transition to a smarter grid is empowered by enhanced sensor deployments and smart metering infrastructure in the distribution system. Measurements from these sensors and meters can be used for many applications, including distribution system state estimation (DSSE). However, these measurements are typically sampled at different rates and could be intermittent due to losses during the aggregation process. These multi time-scale measurements should be reconciled in real-time to perform accurate grid monitoring.  This paper tackles this problem by formulating a recursive multi-task Gaussian process (RGP-G) approach that sequentially aggregates sensor measurements. Specifically, we formulate a recursive multi-task GP with and without network connectivity information to reconcile the multi time-scale measurements in distribution systems. The proposed framework is capable of aggregating the multi-time scale measurements batch-wise or in real-time. 
Following the aggregation of the multi time-scale measurements, the spatial states of the consistent time-series are estimated using matrix completion based DSSE approach.  Simulation results on IEEE 37 and IEEE 123 bus test systems illustrate the efficiency of the proposed methods from the standpoint of both multi time-scale data aggregation and DSSE.
\end{abstract}


\section{Introduction}
Distribution system state estimation (DSSE) techniques infer the system states based on the network model and available measurements. The distribution system typically has limited number of measurement devices to monitor the medium and low-voltage feeders, rendering the system unobservable \cite{abur2004power}, \cite{dehghanpour2018survey}. The lack of  measurement data hinders the development and use of DSSE. In recent years, the installation of different measurement sensors has increased significantly. For example, smart meters are being deployed in large numbers at the secondary side of the distribution systems. They are typically sampled at 15-min intervals and used for consumer billing purposes. The load composition of a primary feeder can be calculated according to the energy consumption of all the customers served by the feeder \cite{373946}. The aggregated smart meter measurements at the primary side are critical measurements for increasing the data redundancy in the distribution system \cite{gomez2012state}. There has also been an increase in the deployment of PMU (Phasor Measurement units) and SCADA (supervisory control and data acquisition) sensors. In addition to these sensors and meters, monitoring data from distributed generation (DG) devices are available periodically. The distribution management system (DMS) also has access to day-ahead forecasting data for load and DG.

\subsection{Problem Statement}
Aggregating the multiple sources of information in a smart grid presents some challenges.  Firstly, the measurements from heterogeneous sources have different sampling rates and are rarely synchronized. The sources of information discussed above can be broadly classified as - (1) Fast rate measurements collected by PMUs or SCADA systems that are typically sampled at rates ranging from few milli-seconds to minute \cite{gomez2014state}, and (2) Slow rate measurements at the primary feeder obtained by smart meter or distribution generation data averaged over 15 minutes or 1 hour. 
 Secondly, the information aggregated from these sources can be intermittent and corrupted due to communication network impairments. 
 DSSE is thought of as a real-time operation. However, the measurements that are received at the DSSE are sampled at different snapshots of time. Also, the AMI measurements are loosely time-synchronized with possible delays of hours
 \cite{feng2012practical}.  Hence, real-time imputation of the slow-rate measurements is necessary for a reliable DSSE
 Finally, it is likely that network topology information available to the utility is incorrect or completely unknown \cite{yu2017patopa}, \cite{karimi2021joint}. Hence, one of the critical challenges in distribution system state estimation is properly aggregating and reconciling noisy, corrupted, heterogeneous, and incomplete time-series data and network topology information for a reliable DSSE. 
 \subsection{Related Work and limitations}
 Previous research efforts have focused on reconciling two time-scale measurements using linear interpolation/extrapolation based weighted least squares (WLS) approach \cite{gomez2014state}. However, this approach does not exploit any underlying spatio-temporal relationships in the time-series data. Authors in \cite{alimardani2015distribution} address the asynchronicity problem of smart meter measurements for DSSE. An extended Kalman filter approach was proposed in \cite{stankovic2017hybrid} to deal with the issue of irregular sensor sampling. A multi-task Gaussian process framework to reconcile heterogeneous measurements was proposed in \cite{9637824}, \cite{dahale2022bayesian}.
  The multitask GP approach proposed in \cite{9637824}, \cite{dahale2022bayesian} performs imputations using all the measurements at once. That is, the approach proposed in \cite{9637824}, \cite{dahale2022bayesian} involves batch processing and cannot be used to perform imputations in real-time as measurements arrive.  Furthermore, these methods do not exploit the graphical structure of the grid.

Recently, sparsity-aware DSSE approaches are proposed to address the issue of low observability at the grid edge \cite{shafiul, liu2019robust, madbhavi2020tensor, joshi2018framework, dahalejoint}. The compressive sensing-based approach estimates the states that are sparse in a linear transformation basis \cite{shafiul}. Matrix completion based DSSE approach exploits the sparsity of spatial states by suitable low-rank approximation \cite{donti2019matrix},\cite{sagan2021decentralized}. Tensor completion fills the missing elements in a tensor by exploiting the spatio-temporal correlation of the measurements \cite{madbhavi2020enhanced}. A comparative analysis of these sparsity-based approaches along with their robust formulations was proposed in \cite{9247106}.
 Authors in \cite{alcaide2017electric} use PMU and SCADA measurements for DSSE. This approach performs DSSE by incorporating a subset of these measurements available at time $t$ along with the predicted SCADA measurements obtained using the information from the previous state estimates. It suffers from large measurement redundancy requirements (around 1.7), which makes it impractical for low-observable distribution systems. Furthermore, \cite{alcaide2017electric} does not consider any missing measurements scenario that could occur while aggregating measurements over finite bandwidth communication networks. A load evolution model for the slow-rate measurements is proposed in \cite{carquex2018state} for performing day-ahead forecasting. This approach relies on recursive Kalman filter (KF) updates for dynamic DSSE. However, KF typically needs the Hessian inverse computations at every step, which can be computationally burdensome. Also, the approach is not demonstrated for unbalanced systems. A first-order prediction-correction approach using PMU and smart meter data is performed in \cite{song2019dynamic}. The main limitation of the approaches in \cite{carquex2018state} and \cite{song2019dynamic} approaches is that they assume smart meter measurements are available at all load bus (i.e., system is fully observable).
 
This paper proposes a recursive multi-task Gaussian process approach that sequentially aggregates multi-time scale measurements depending on the network connectivity information. It addresses multiple limitations of the state-of-the-art approaches. For example, the proposed approach is: - (1) Flexible to incorporate various heterogeneous measurements for unbalanced systems; (2) Does not require power measurements at all the load buses. The considered measurement dataset represents an unobservable
condition that is used for both imputation and DSSE; (3) Effective even with missing data within each measurement time-series, and (4) Since inversion computations are only required at the initial time step, the proposed approach is computationally efficient. 
 \subsection{Contributions}
 The main contributions of this paper are summarized below:
 \begin{itemize}
     \item We propose a novel approach that imputes the heterogeneous measurements sequentially at any desired time resolution using recursive multi-task GP with or without topology information.
     
     \item The proposed approach involves sequential measurement processing and can work with intermittent measurements. Unlike the approach in \cite{gomez2014state}, the proposed method is computationally efficient and is flexible to allow for both batch-wise and real time processing of measurements.

     \item Finally, we leverage the graphical structure of the network in the recursive multi-task Gaussian process approach. We prove that exploiting the graph structure of the distribution system leads to a decrease in the variance of the imputed measurements. 
     
     \item Simulation results are carried out for the IEEE 37 and IEEE 123 bus test systems to verify the efficacy of the proposed approach. Relative to the linear interpolation approach \cite{gomez2014state}, the RGP-G approach offers nearly 80\% improvement in error performance while  reconciling the multi time-scale measurements. We further estimate the spatial states of the consistent time-series measurements using the matrix completion based DSSE proposed in \cite{donti2019matrix}. It can be inferred that accurate reconstruction of states is achieved even at 50\% FAD (fraction of available data, which reflects the number of available measurements in the system).
     
 \end{itemize}
\section{Background}

Consider a distribution system which can be perceived as a graph $\mathbf{\mathcal{G} = ({\mathcal{V},\mathcal{E}})}$ where $\mathcal{V} \in \mathbb{R}^{M}$ are the nodes and $\mathcal{E}$ denotes the edges.  The adjacency matrix $\mathbf{A}$ is defined as,
\begin{align}
  \mathbf{A}(i,j) = 
  \begin{cases}
  1 , \enspace (i,j) \in \mathcal{E}
  \\
  0, \enspace \enspace\text{otherwise}
  \end{cases}
\end{align}    
In the distribution system, sensors are placed at a subset of $M$ nodes. These sensors measure power injections or voltages at different locations in the network at different sampling rates. Hence, the main goal is to first reconcile these multi time-series measurements at the desired time scale and estimate the states. To do so, we sequentially process the measurements using a recursive multi-task Gaussian process-aided state estimation approach. This approach exploits the spatial and temporal correlations of the sequential measurements located on the graph $\mathcal{G}$. We propose to leverage the graphical structure of the grid for better imputation of the unevenly sampled measurements. Here, we consider the topology of the primary distribution system. Also, the imputation of the multi time-scale measurements is performed at the primary feeder. Therefore, the proposed approach does not rely on the model information of secondary side of the network. Based on the available network connectivity information, we propose two approaches, as shown in Fig.1. The inputs in all these approaches are the unevenly sampled time-series measurements, and output is the coherent set of measurements along with their variance. Conventional full GP and multitask Recursive GP (RGP) process the unevenly sampled measurements without utilizing any graph structure. Recursive processing of heterogeneous measurements using graph information in real-time and batch mode is performed using RGP-G interpolation and RGP-G prediction, respectively.  We will review the relevant concepts of graph signal processing before introducing the proposed approaches. 
\begin{figure}[h!]
\centering
\includegraphics[width= 0.3\textwidth]{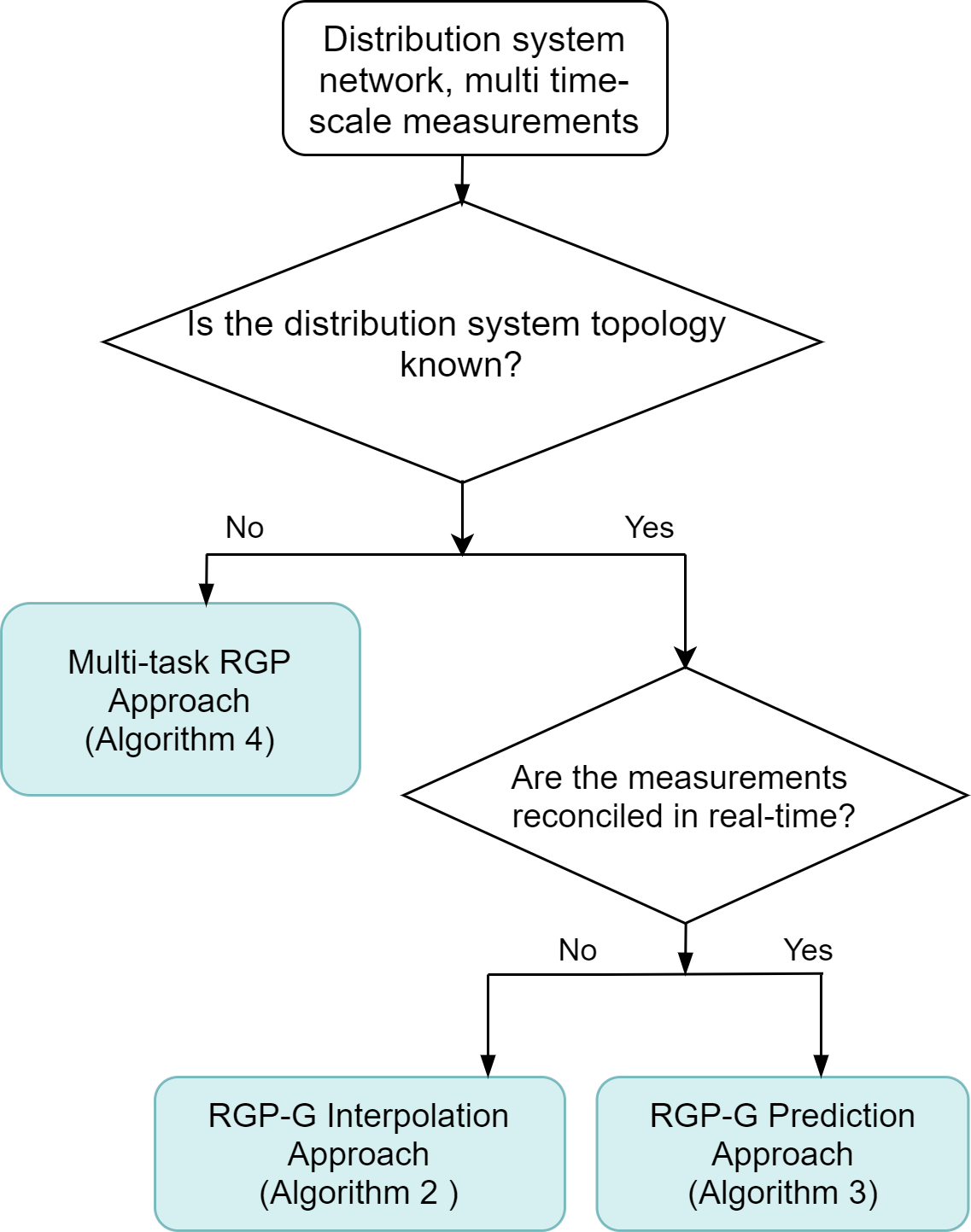}
\caption{{Classification of the proposed approaches}}
\label{fig:vtgmag_recovery2}
\end{figure}
\subsection*{Background of Graph signal processing}
The graph Laplacian matrix $\mathbf{L}$ for graph $\mathcal{G}$  is defined as $\mathbf{L = D-A}$ where $\mathbf{D}$ is the diagonal degree matrix whose $i$th diagonal element is given by the sum of the elements in the $i$th row of $\mathbf{A}$.
   The observations $\mathbf{y} = [y(1),...,y(M)] \in \mathbb{R}^M$ represents a signal on graph $\mathcal{G}$. The signal variation of $\mathbf{y}$ over graph $\mathcal{G}$ is measured as,
   \begin{equation*}
   l(\mathbf{y}) = \sum_{(i,j) \in \mathcal{E}, i \neq j}A(i,j) ((y(i)- y(j))^2 = \mathbf{y^\intercal L y}
   \end{equation*}
The Laplacian quadratic form $\mathbf{y^\intercal L y}$ denotes the smoothness of the  $\mathbf{y}$. 
  Suppose we want to recover the smooth signal $\mathbf{y_d}$ from noisy observation $\mathbf{y = y_d + w}$, over the graph $\mathcal{G}$. In order to recover this signal, an optimization problem is formulated as,
\begin{equation}
\begin{aligned}
{\mathbf{y_d}^*} = & \underset{\mathbf{y_d}  \in \mathbb{R}^M}{\text{\enspace min}} \enspace
 \| \mathbf{y - y_d}\|^2_2 + \alpha \mathbf{y_d^\intercal L y_d}  \\
\end{aligned}
\label{eq14}
\end{equation}
where $\alpha \geq 0$. The global solution is,
\begin{equation}
    \mathbf{y_d}^* = \mathbf{(I}_M + \alpha \mathbf{L})^{-1} \mathbf{y} 
    \label{graph_filter}
\end{equation}
Here, $\mathbf{I}_M$ is the identity matrix.  The optimal solution $\mathbf{y_d^*}$ can be seen as the graph filtering of $\mathbf{y}$ using the  graph filter $\mathbf{S} =  \mathbf{(I}_M + \alpha \mathbf{L})^{-1}$ \cite{venkitaraman2020gaussian}\cite{shuman2013emerging}. This graph filter will be used in the proposed RGP-G approach for inducing the graph structure of the distribution system. In the next section, we will review the conventional full GP approach.
\subsection{Full GP Approach}
Consider a distribution system with $M$ buses and $d$ types of sensor tasks. Here, the sensor tasks refer to different sensor measurements available, e.g., aggregated active and reactive power injections or voltage magnitudes at the primary feeder. We consider the availability of measurements for $T$ time instances.  Let the measurements corresponding to time instant $x_t$ be $\mathbf{y}_t \in \mathbb{R}^{dM}$. The measurements $\mathbf{y}_t$ are obtained by concatenating measurements from different sensor locations i.e., $\mathbf{y}_t  = [\mathbf{y}^{1}_t,..., \mathbf{y}^{d}_t]^\intercal$. 
Each $\mathbf{y}^d_t$ is drawn from a noisy process as,
\begin{equation}
  \mathbf{y}^d_t(x_t) = \mathcal{N} (\mathbf{f}^d_t (x_t),\sigma_\epsilon^2  \mathbf{I}_M)
  \label{observation_GP}
\end{equation}
where, $\mathbf{f}^d_t \in \mathbb{R}^{M}$ and $\sigma_\epsilon^2$ is the noise variance. The entries in $\mathbf{y}^d_t$ are zero at the locations where there are no sensor measurements.

The GP prior function $\mathbf{f}^d_t$ associated with $d^{th}$ sensor task at time $t$ has distribution given as,
\begin{equation}
   \mathbf{f}^d_t = \mathcal{N}(\mathbf{0}, \mathcal{K}(x_t,x_t^\intercal)\mathbf{I}_M)
   \label{ftd}
\end{equation}

The function $\mathbf{f}_t = [\mathbf{f}^1_t,...,\mathbf{f}^d_t]^\intercal$ is a Gaussian prior with distribution,
\begin{equation}
  \mathbf{f}_t(x_t) = \mathcal{N} (\mathbf{0}, \mathbf{I} _{c} \otimes \mathcal{K}(x_t, x_t^\intercal)\mathbf{I}_M)
  \label{gp_prior_MOGP}
\end{equation}
where $\otimes$ denotes the Kronecker product, $\mathbf{I} _{c} \in \mathbb{R}^{d \times d}$ is an identity matrix between different sensor tasks in the distribution grid. For instance, a distribution grid may have active power ($\mathbf{P}$), reactive power injections ($\mathbf{Q}$), and voltage ($\mathbf{V}$) measurements at the primary feeder. Thus, there are three sensors tasks, and $\mathbf{I}_c$ has a size of $3 \times 3$. In some distribution systems, there are only $\mathbf{P}$ and $\mathbf{Q}$ measurements available. The voltage  measurements are available only at the substation. Therefore, in this case, the size of $\mathbf{I}_c$ is $2 \times 2$. The kernel matrix  $\mathcal{K}$ represents the temporal covariance functions within this sensor task. There are different kernel choices, with one of the most popular being RBF (radial basis function) kernel \cite{rasmussen2003gaussian} corresponding to:
\begin{equation}
    \mathcal{K}(x_1,x_2) = \sigma_s^2 \text{exp}\frac{-(x_1-x_2)^2}{2l^2} 
    \label{hyperparameter}
\end{equation}
where hyperparameters $l$ and $\sigma^2_s$ are the length-scale and
signal variance respectively. The lengthscale of the kernel function controls the smoothness of the GP function \cite{rasmussen2003gaussian}.

If all the measurements upto time $T$ are represented as $\mathbf{\Tilde{y}} = vec(\mathbf{y}_1,...,\mathbf{y}_T)$, the distribution of $ \mathbf{\Tilde{y}}$ using (\ref{observation_GP}) and (\ref{gp_prior_MOGP}) is given as,
\begin{equation}
    \mathbf{\Tilde{y}} = \mathcal{N} ( \mathbf{0}, ((\mathbf{I}_{c} \otimes \mathbf{I}) \otimes \mathbf{K}) + \sigma_\epsilon^2  \mathbf{I} )
    \label{MOGP}
\end{equation}
where $\mathbf{K}$ is the kernel matrix defined for all time instances $t = 1,...,T$. Here, the $i^{th}$ and $j^{th}$ entry of $\mathbf{K}$ is given as $\mathbf{K}_{ij} = \mathcal{K}(x_i, x_j)$ and $\sigma_\epsilon^2$ is the noise variance.

The main goal of the Gaussian process-based imputation process is to infer the unknown test values $\mathbf{y_*}$ corresponding to the time $\mathbf{x}_*$ given the measurements $\mathbf{\Tilde{y}}$ at time $\mathbf{x}$ and the modeled GP prior function $\mathbf{f}(\cdot)$.
The measurements $ \mathbf{\Tilde{y}}$ and the test values $\mathbf{y_*}$ are jointly Gaussian whose distribution is given as,
\begin{eqnarray}
\begin{pmatrix}\mathbf{\Tilde{y}}\\
\mathbf{y}_*\\
\end{pmatrix} & \sim & \mathcal{N} \left(\mathbf{0},\begin{pmatrix}
\mathbf{A} & \mathbf{D} \\
\mathbf{D}^\intercal & \mathbf{F}
\end{pmatrix} \right)
\label{eqn11}
\end{eqnarray}
where, the matrices $\mathbf{A}$, $\mathbf{D}$ and  $\mathbf{F}$ corresponds to,
\begin{equation}
\mathbf{A} = (\mathbf{I}_{c} \otimes \mathbf{I}) \otimes \mathbf{K} + \sigma_{\epsilon}^2 \mathbf{I}, 
\label{A_fullGP}
\end{equation}
 \begin{equation}
\mathbf{D} = (\mathbf{I}_{c} \otimes \mathbf{I}) \otimes \mathbf{K}_* + \sigma_{\epsilon}^2 \mathbf{I}, 
\label{D_fullGP}
\end{equation}
\begin{equation}
\mathbf{F} = (\mathbf{I}_{c} \otimes \mathbf{I}) \otimes \mathbf{K}_{**} + \sigma_{\epsilon}^2 \mathbf{I}.
\label{F_fullGP}
\end{equation}
Here,  $\mathbf{K}_* = \mathcal{K}\left(\mathbf{x}, \mathbf{x}_*\right)$, $\mathbf{K}_{**} = \mathcal{K}\left(\mathbf{x}_*, \mathbf{x}_*\right)$ and $\mathbf{x}= [x_1, ..., x_T]^\intercal $. 

The conditional distribution of the test values  $\mathbf{y_*}$ given $\mathbf{\Tilde{y}}$  is a Gaussian distribution \cite{rasmussen2003gaussian} with mean and covariance, 
\begin{equation}
    \mathbf{m^*} = \mathbf{D}^\intercal \mathbf{A}^{-1} \mathbf{\Tilde{y}}
    \label{fullgp_m*}
\end{equation}
and
\begin{equation}
    \mathbf{C^*} = \mathbf{F} - \mathbf{D}^\intercal \mathbf{A}^{-1} \mathbf{D}
    \label{fullgp_C*}
\end{equation}

Algorithm 1 summarizes the full-GP approach. (\ref{fullgp_m*}) and (\ref{fullgp_C*}) involves inverting the matrix $\mathbf{A}$ for all the $\mathbf{x}$ time instances which is  computationally expensive. 
The full GP approach suffers from the following drawbacks:
\begin{itemize}
    \item The GP prior function and the corresponding measurements $\mathbf{\Tilde{y}}$ as defined in (\ref{MOGP}) is a simple multi-task Gaussian process with an independent kernel function among the different measurements obtained at $M$ nodes. 
    \item This approach performs training using all the measurements in the batch, and thus the training is performed off-line in a batch mode.
    \item The computational complexity is $\mathcal{O}((TdM)^3)$ 
    ,where $T$ is the size of $\mathbf{x}$, $d$ is the total number of sensor tasks, and $M$ are the nodes. The inversion of the matrix $\mathbf{A}$ is the key contributor to this complexity. 
\end{itemize}
To overcome these challenges, we propose a recursive GP approach that sequentially processes the measurements corresponding to each $x_t$ by using the knowledge of graphical structure of the distribution grid. In the next section, we will formulate the RGP-G approach when the network connectivity information is known. Then, we will develop the RGP-G method when the graph information is unknown. 

\begin{algorithm}
\KwInput{Aggregated Active and Reactive power injection  measurements at load bus $\mathbf{y}_t$ corresponding to time $\mathbf{x}_t$,  $\mathbf{\Tilde{y}} = vec(\mathbf{y}_1,...,\mathbf{y}_T)$, Kernel choice and hyper-parameters associated to kernel function}
\begin{algorithmic}[1]
 
\STATE Calculate the kernel matrix $\mathbf{K}$ that exploits the temporal correlation using any kernel function (e.g., RBF kernel).

\STATE Calculate the matrix $\mathbf{I}_{c} \in \mathbb{R}^{d \times d}$.

\STATE Calculate the matrix $\mathbf{A}$, $\mathbf{D}$ and $\mathbf{F}$ by means of (\ref{A_fullGP}), (\ref{D_fullGP}) and (\ref{F_fullGP}) respectively.

\STATE Perform imputation at time $x_*$ by means of the mean $\mathbf{m}^*$ and covariance matrix $\mathbf{C}^*$ using (\ref{fullgp_m*}) and (\ref{fullgp_C*}).  
    \STATE \textbf{return} $\mathbf{m}^*$, $\mathbf{C}^*$.
  \end{algorithmic}
  \caption{Full GP Approach}
\end{algorithm}

\section{Proposed Approach}
This section presents the formulation for recursively imputing the multi time-scale measurements with and without topology information.
\subsection{RGP-G Approach}
One of the challenges in the full GP approach is the need to to use the complete vector  $\mathbf{x} \in \mathbb{R}^T$. To overcome this challenge, we aim to use the basis vectors $\mathbf{x} = [x_1, x_2, ..., x_n]^\intercal$ where $n \ll T$. We perform all the calculations on the basis vectors $\mathbf{x} \in \mathbb{R}^n$ which are fixed in number and locations. The function $\mathbf{f} = \mathbf{f(x)}$ is the GP function corresponding to the basis vectors $\mathbf{x}$. When the network connectivity information is known, we can construct the graph filter matrix $\mathbf{S}$  as defined in (\ref{graph_filter}). In order to exploit the spatial correlation and induce the graph information, the observations in the GP function (4) and (5) are modified as,
\begin{equation}
  \mathbf{y}^d_t(x_t) = \mathcal{N} (\mathbf{S} \mathbf{f}^d_t (x_t),\sigma_\epsilon^2  \mathbf{I}_M)
  \label{ytd_rgpg}
\end{equation}
where, $\mathbf{f}^d_t$ is defined in (\ref{ftd}). Using (\ref{ftd}) and (\ref{ytd_rgpg}), we obtain the distribution of $\mathbf{y}^d_t$ as,
\begin{equation}
 \mathbf{y}^d_t = \mathcal{N} (\mathbf{0}, \mathbf{S} \mathcal{K}(x,x^\intercal) \mathbf{S}^\intercal +\sigma_\epsilon^2  \mathbf{I}_M)
\end{equation}
The distribution of $\mathbf{y}_t$ obtained by concatenating $\mathbf{y}^d_t$ is given as, 
\begin{equation}
    \mathbf{y}_t(x_t) = \mathcal{N} ( \mathbf{0}, (\mathbf{K}_{c} \otimes \mathbf{S}^2) \mathcal{K}(x_t,x_t') + \sigma_\epsilon^2  \mathbf{I}  )
\end{equation}
Here, the identity matrix $\mathbf{I}_c$ given in (\ref{gp_prior_MOGP}) is replaced by the kernel matrix $\mathbf{K}_c$ that represents correlation among different sensor tasks. 

The main aim of this section is to recursively update the mean and covariance of the multi-task prior function $\mathbf{f}$ as the measurements arrive at time $t = 1,..., T$ by incorporating the graph structure of the grid. We assume that the hyperparameters of the kernel function are known apriori using historical-based data. The proposed RGP-G approach has the flexibility of performing both interpolation and prediction described as, 



\begin{enumerate}
\item  RGP-G Interpolation- This approach operates over a set time frame (24 hours as an example). The GP function is updated recursively at those time instances where the measurements are obtained.    Once the GP function in that batch is updated, we perform imputation at the finest time resolution. Here, the finest time resolution refers to the narrowest time resolution between the different measurement sources.


\item  RGP-G Prediction-
This approach reconciles the multi-time scale measurements in real-time. The imputation at the desired time resolution is performed by predicting the GP function until the subsequent measurements is observed. Here, the  prediction is performed at the finest time resolution.
\end{enumerate}
   
\subsubsection{RGP-G Interpolation}  
    We assume that the GP prior function $\mathbf{f}$ at time $t=0$ has an initial distribution,
    \begin{equation}
    p_0(\mathbf{\mathbf{f}}) = \mathcal{N} (\mathbf{f};\boldsymbol{\mu}_{g,0}^\mathbf{f}, \mathbf{C}_{g,0}^\mathbf{f})
    \label{initial_RGPG}
    \end{equation}
with mean $\boldsymbol{\mu}_{g,0}^\mathbf{f} = \mathbf{0}$ and covariance $\mathbf{C}_{g,0}^\mathbf{f}$ defined as,
\begin{equation}
\mathbf{C}_{g,0}^\mathbf{f} = (\mathbf{K}_{c} \otimes \mathbf{S}^2) \otimes \mathbf{K}.
\label{Cg0}
\end{equation}
Here, $\mathbf{K} = \mathcal{K}(\mathbf{x,x})$. The initial covariance of the GP prior function exploits the spatial correlation using graph structure and the temporal correlation between the time instances. 
    The measurements $\mathbf{y}_t \in \mathbb{R}^{dM}$ arrive sequentially at time $t  = 1,..,T$. The goal is to calculate the posterior distribution
    \begin{equation}
       p(\mathbf{f}|\mathbf{y}_{1:t}) = \mathcal{N}(\mathbf{f}; \boldsymbol{\mu}_{g,t}^\mathbf{f}, \mathbf{C}_{g,t}^\mathbf{f})
    \end{equation} 
    at time $t$, where $\mathbf{y}_{1:t} = (\mathbf{y}_1, ..., \mathbf{y}_t)$, by combining the new measurements $\mathbf{y}_t$ with the distribution,
    \begin{equation}
        p_{t-1}(\mathbf{f}|\mathbf{y}_{1:t-1}) = \mathcal{N}(\mathbf{f}; \boldsymbol{\mu}_{g,t-1}^\mathbf{f}, \mathbf{C}_{g,t-1}^\mathbf{f})
    \end{equation}
   
The desired posterior distribution is expanded according to \cite{huber2014recursive},
\[ p(\mathbf{f|y}_{1:t}) = \int \underbrace{c_t \cdot p(\mathbf{y}_t| \mathbf{f},\mathbf{f}_t) \cdot \overbrace{p(\mathbf{f}_t|\mathbf{f}) \cdot p(\mathbf{f}|\mathbf{y}_{1:t-1})}^{p(\mathbf{f},\mathbf{f}_t|\mathbf{y}_{1:t-1}) \enspace \text{inference}}   }_{p(\mathbf{f},\mathbf{f}_t| \mathbf{y}_{1:t}) \enspace \text{update}} d\mathbf{f}_t\]
 where, $\mathbf{f}_t$ is the GP function at time $t$ and $c_t$ is the normalization constant.

Calculation of the posterior is performed in two steps:\\
a) Inference: In this step, we infer the joint prior $p(\mathbf{f},\mathbf{f}_t| \mathbf{y}_{1:t-1})$  using the measurements received upto time $t-1$. Here, the matrices $\mathbf{A}_g,\mathbf{D}_g, \mathbf{F}_g$ calculated at time $x_t$ be defined as, 
\begin{equation}
\mathbf{A}_g = (\mathbf{K}_{c} \otimes \mathbf{S}^2) \otimes \mathbf{K} + \sigma_{\epsilon}^2 \mathbf{I},
\label{A_RGPG}
\end{equation}
\begin{equation}
\mathbf{D}_g = (\mathbf{K}_{c} \otimes \mathbf{S}^2) \otimes \mathbf{K}_t + \sigma_{\epsilon}^2 \mathbf{I},
\label{D_RGPG}
\end{equation}
\begin{equation}
\mathbf{F}_g = (\mathbf{K}_{c} \otimes \mathbf{S}^2) \otimes \mathcal{K}_{tt} + \sigma_{\epsilon}^2 \mathbf{I}
\label{F_RGPG}
\end{equation}
Here, the subscript $g$ refers to notations related to recursive GP with graphs approach. 
 The goal is to calculate the joint prior $p(\mathbf{f},\mathbf{f}_t| \mathbf{y}_{1:t-1})$ using the information from the prior $p({\mathbf{f}|\mathbf{y}_{1:t-1}})$. This can be achieved using the chain rule as, 
    \begin{align}
        p(\mathbf{f},\mathbf{f}_t| \mathbf{y}_{1:t-1}) &= p(\mathbf{f}_t|\mathbf{f}) \cdot p(\mathbf{f}|\mathbf{y}_{1:t-1}) \\
        &= \mathcal{N}(\mathbf{f}_t;\boldsymbol{\mu}_{g,t}^\mathbf{p}, \mathbf{B}_g) \cdot \mathcal{N}(\mathbf{f};\boldsymbol{\mu}_{g,t-1}^\mathbf{f}, \mathbf{C}_{g,t-1}^\mathbf{f})
        \label{eqn19}
    \end{align}
The first term  $p(\mathbf{f}_t|\mathbf{f})$ 
follows from the assumption that $\mathbf{f}_t$ is conditionally independent of the past measurements $\mathbf{y}_{1:t-1}$ given $\mathbf{f}$. As any finite representation of a GP is Gaussian, the joint prior is also Gaussian. Hence, the conditional distribution $p(\mathbf{f}_t|\mathbf{f})$ is Gaussian and calculated by Gaussian identities given as,
\begin{equation}
\boldsymbol{\mu}_{g,t}^\mathbf{p}= \mathbf{J}_{g,t} \boldsymbol{\mu}_{g,t-1}^\mathbf{f}
\label{mug_RGPG}
\end{equation}    
\begin{equation}
\mathbf{B}_g = \mathbf{F}_g - \mathbf{J}_{g,t} \mathbf{D}_g
\label{B_g_RGPG}
\end{equation}
\begin{equation}
\mathbf{J}_{g,t} = \mathbf{D}_g^\intercal \mathbf{A}_g^{-1}
\label{J_RGPG}
\end{equation}

Using Gaussian identities and Woodbury formula, the solution to (\ref{eqn19}) is a joint Gaussian $p(\mathbf{f},\mathbf{f}_t| \mathbf{y}_{1:t-1}) = \mathcal{N}(\mathbf{q,Q})$ with mean and covariance defined as, 
\begin{equation}
\mathbf{q}  =  
  \begin{bmatrix}
   \boldsymbol{\mu}_{g,t}^\mathbf{f} \\[4pt] \boldsymbol{\mu}_{g,t}^\mathbf{p}
  \end{bmatrix}
\end{equation}
and,
\begin{equation}
\mathbf{Q}  =  
  \begin{bmatrix}
   \mathbf{C}_{g,t}^\mathbf{f}  & \mathbf{C}_{g,t-1}^\mathbf{f} \mathbf{J}_{g,t}^\intercal \\
   \mathbf{J}_{g,t} \mathbf{C}_{g,t-1}^\mathbf{f} & \mathbf{C}_{g,t}^\mathbf{p}
  \end{bmatrix}
\end{equation}
where, 
\begin{equation}
\mathbf{C}_{g,t}^\mathbf{p} = \mathbf{B}_g + \mathbf{J}_{g,t} \mathbf{C}_{g,t-1}^\mathbf{f} \mathbf{J}_{g,t}^\intercal
\label{cg_RGPG}
\end{equation}
    b) Update: This step updates the joint prior $\mathbf{f}$ with new measurements $\mathbf{y}_t$ arriving at time $t$. The function $\mathbf{f}_t$ is updated by Kalman filter update step which yields, 
    \begin{equation}
        p(\mathbf{f}_t|\mathbf{y}_{1:t}) = \mathcal{N}(\mathbf{f}_t;\boldsymbol{\mu}_{g,t}^\mathbf{e}, \mathbf{C}_{g,t}^\mathbf{e})
    \end{equation}
where,
\begin{equation}
    \boldsymbol{\mu}_{g,t}^\mathbf{e} =  \boldsymbol{\mu}_{g,t}^\mathbf{p} + \mathbf{G}_t (\mathbf{y}_t -  \boldsymbol{\mu}_{g,t}^\mathbf{p}),
\end{equation}
\begin{equation}
    \mathbf{C}_{g,t}^\mathbf{e} =  \mathbf{C}_{g,t}^\mathbf{p} - \mathbf{G}_t (\mathbf{C}_{g,t}^\mathbf{p}),
\end{equation}    
Here, $\boldsymbol{\mu}_{g,t}^\mathbf{p}$ and $\mathbf{C}_{g,t}^\mathbf{p}$  are obtained from (\ref{mug_RGPG}) and (\ref{cg_RGPG}), respectively.
Here, $\mathbf{G}_t = \mathbf{C}_{g,t}^\mathbf{p} (\mathbf{C}_{g,t}^\mathbf{p} + \sigma_{\epsilon}^2 \mathbf{I})^{-1}$ is the Kalman gain. 
The update is performed at time $t$ where measurements are available. For instance, the aggregated smart meter at the load buses provides measurements at intervals $t = 1, 16, 31,..., T$. At $t = 16$, assume that few of the measurements $\mathbf{y}_t$ are missing due to communication bottleneck. Therefore, at those locations, we do not update but use the predicted measurements $\boldsymbol{\mu}_{g,t}^\mathbf{p}$. The posterior function $\mathbf{f}$ has Gaussian distribution $\mathcal{N}(\mathbf{f};\boldsymbol{\mu}_{g,t}^\mathbf{f}, \mathbf{C}_{g,t}^\mathbf{f})$ which is defined as,
\begin{equation}
\boldsymbol{\mu}_{g,t}^\mathbf{f} = \boldsymbol{\mu}_{g,t-1}^\mathbf{f} + \Tilde{\mathbf{G}}_t \cdot (\mathbf{y}_t - \boldsymbol{\mu}_{g,t}^\mathbf{p}),
\label{mugt_RGPG}
\end{equation}
\begin{equation}
\mathbf{C}_{g,t}^\mathbf{f} =\mathbf{C}_{g,t-1}^\mathbf{f} - \Tilde{\mathbf{G}}_t \mathbf{J}_{g,t} \mathbf{C}_{g,t-1}^\mathbf{f},
\label{Cgt_RGPG}
\end{equation}
\begin{equation}
\Tilde{\mathbf{G}}_t =\mathbf{C}_{g,t-1}^\mathbf{f} \mathbf{J}_{g,t}^\intercal (\mathbf{C}_{g,t}^\mathbf{p} +\sigma_{\epsilon}^2 \mathbf{I})^{-1}.
\label{Ghat_RGPG}
\end{equation}

The function $\mathbf{f}$ is sequentially updated with the observations $\mathbf{y}_t$ until $t=T$. Once updated, the imputation of $\mathbf{y}_*$ at time $x_*$ is performed using the following steps, 
\begin{equation}
    \mathbf{m^*}_g = \mathbf{D}_g^\intercal \mathbf{A}_g^{-1} \boldsymbol{\mu}_{g,T}^\mathbf{f}
    \label{M*RGPG}
\end{equation}
and
\begin{equation}
    \mathbf{C^*}_g = 
    \mathbf{B}_g +  \mathbf{J}_{g,t^*} (\mathbf{C}_{g,T}^\mathbf{f}) \mathbf{J}_{g,t^*}^\intercal
    \label{C*RGPG}
\end{equation}
where the matrices $\mathbf{D}_g$, $\mathbf{A}_g$ and $\mathbf{B}_g$ are evaluated for time $x_*$. The complete RGP-G Interpolation approach is illustrated in Algorithm 2.

\begin{algorithm}
\KwInput{Basis vector $\mathbf{x}$, Distribution system graph laplacian $\mathbf{L} \in$ $\mathbb{R}^{M \times M}$, total time instants $T$, $\alpha$, $\mathbf{S}$, $\mathbf{K}$, $\mathbf{K}_c$, $\mathbf{y}_t$}
\begin{algorithmic}[1]

\STATE Initialization: $\boldsymbol{\mu}_{g,0}^\mathbf{f} = \mathbf{0}$ and $\mathbf{C}_{g,0}^\mathbf{f} = (\mathbf{K}_{c} \otimes \mathbf{S}^2) \otimes \mathbf{K}$.
  
    \FOR{$t = 1,...,T$}
\STATE  Calculate the gain matrix $\mathbf{J}_{g,t}$ according to (\ref{J_RGPG}). 

\STATE Calculate mean $\boldsymbol{\mu}_{g,t}^\mathbf{p}$ using (\ref{mug_RGPG}) and covariance matrix $\mathbf{C}_{g,t}^\mathbf{p}$ using (\ref{cg_RGPG}).

\STATE Calculate the gain matrix $\Tilde{\mathbf{G}}_t$ according to (\ref{Ghat_RGPG}). 

\STATE Set $\mathbf{y_t} = \boldsymbol{\mu}_{g,t}^\mathbf{p}$ at locations where $\mathbf{y_t}$ is missing.  Calculate mean $\boldsymbol{\mu}_{g,t}^\mathbf{f}$ by means of (\ref{mugt_RGPG}) and covariance matrix $\mathbf{C}_{g,t}^\mathbf{f}$ by means of (\ref{Cgt_RGPG}). 
    \ENDFOR
    \STATE Perform imputation at time $x_*$ by means of mean $\mathbf{m}_g^*$ (\ref{M*RGPG}) and covariance matrix $\mathbf{C}_g^*$ (\ref{C*RGPG}).  
    \STATE \textbf{return} $\mathbf{m}_g^*$, $\mathbf{C}_g^*$.
  \end{algorithmic}
  \caption{RGP-G Interpolation Approach}
\end{algorithm}

\subsubsection{RGP-G Prediction Approach}
 The RGP-G interpolation approach operates  over  a  set  time frame.
 However, it is critical to reconcile the measurements as and when they arrive. The RGP-G prediction approach achieves the reconciliation in real-time by performing the step ahead prediction of the GP function using the knowledge of the past measurements. These predictions are the imputed values at the narrowest time resolution.

\begin{figure*}[h!]
\centering
\includegraphics[width= 0.75\textwidth]{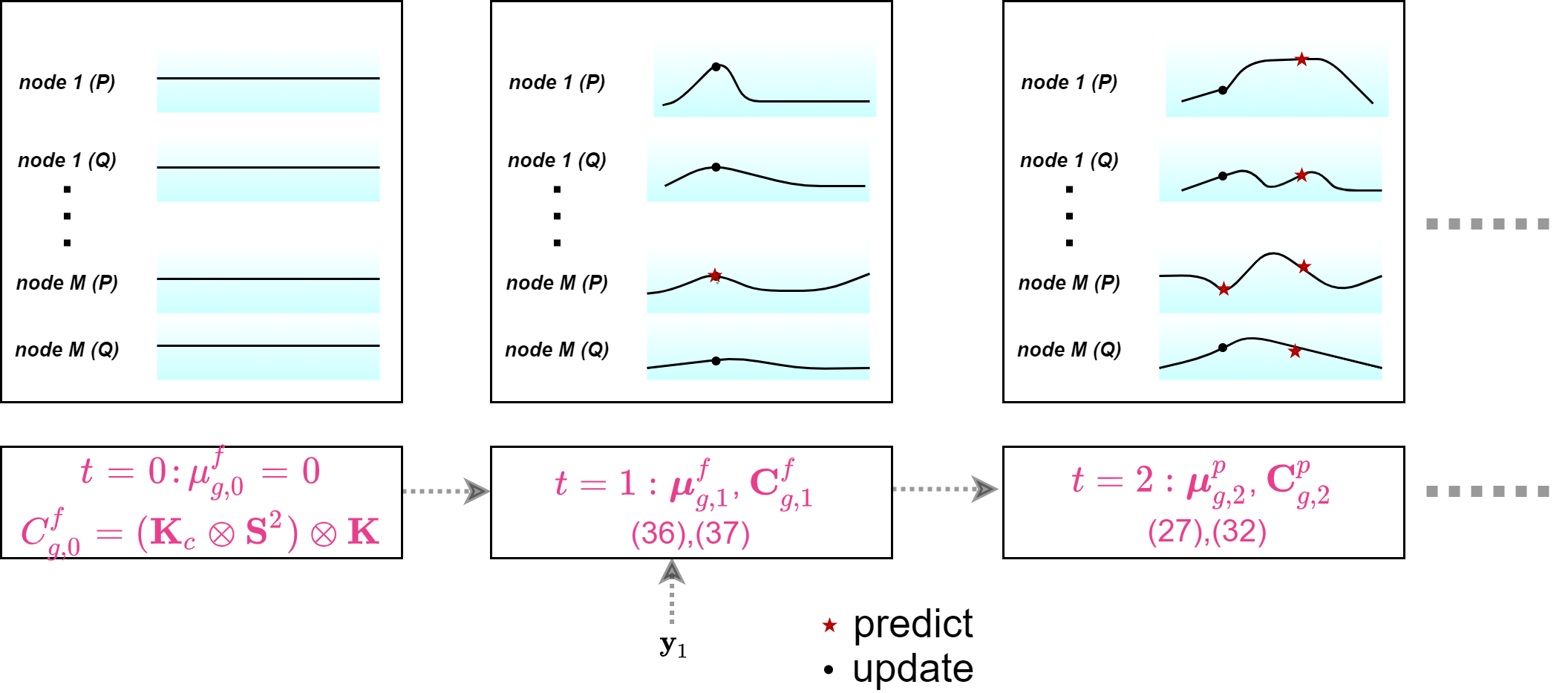}
   \caption{{RGP-G prediction approach}}
     \label{fig:rgpg_flow}
 \end{figure*}

 The complete algorithm of the proposed sequential prediction over graphs is summarized in Algorithm 3. Fig.\ref{fig:rgpg_flow} shows the proposed RGP-G prediction approach. We demonstrate this approach by illustrating an example. 
 At time $t = 0$, we initialize the GP function as given in (\ref{initial_RGPG}). At time $t = 1$, we receive measurements $\mathbf{y}_1$. The GP function is updated using these measurements $\mathbf{y}_1$ by means of (\ref{mugt_RGPG}) and (\ref{Cgt_RGPG}). The updation of the mean and covariance of the GP function $\mathbf{f}$ are denoted by $\boldsymbol{\mu}_{g,t}^\mathbf{f}$ and $\mathbf{C}^\mathbf{f}_{g,t}$. If any measurements at time $t=1$ are missing, they are predicted using (\ref{mug_RGPG}) and (\ref{cg_RGPG}). From time $t=2$ onwards, we perform step-ahead prediction of the GP function using (\ref{m*_sequential}) and (\ref{C*_sequential}) until the subsequent measurements are observed. We perform prediction for all $x(t_*)$ satisfying $x(t_*) > x(t)$ using the knowledge of function $\mathbf{f}$ updated at the previous time instant. The predicted mean $\mathbf{m}_g^*$ and their associated variances $\mathbf{C}_g^*$ are given as,
\begin{equation}
  \mathbf{m}_g^* = \mathbf{D}_g^\intercal \mathbf{A}_g^{-1} \boldsymbol{\mu}_{g,t}^\mathbf{f}
  \label{m*_sequential}
\end{equation}
and
\begin{equation}
    \mathbf{C}_g^* = 
    \mathbf{B}_g +  \mathbf{J}_{g,t*} (\mathbf{C}^\mathbf{f}_{g,t}) \mathbf{J}_{g,t*}^\intercal
     \label{C*_sequential}
\end{equation}

\begin{algorithm}
\KwInput{Basis vector $\mathbf{x}$, Distribution graph $\mathbf{L} \in$ $\mathbb{R}^{M \times M}$, total time instants $T$, $\alpha$,  $\mathbf{y}_t$, $\mathbf{S}$, $\mathbf{K}$, $\mathbf{K}_c$ }
\begin{algorithmic}[1]

\STATE Set $\boldsymbol{\mu}_{g,0}^\mathbf{f} = \mathbf{0}$ and $\mathbf{C}_{g,0}^\mathbf{f} = (\mathbf{K}_{c} \otimes \mathbf{S}^2) \otimes \mathbf{K}$.
  
\STATE  Calculate the gain matrix $\mathbf{J}_{g,t}$ according to (\ref{J_RGPG}). 

\STATE Calculate mean $\boldsymbol{\mu}_{g,t}^\mathbf{p}$ using (\ref{mug_RGPG}) and covariance matrix $\mathbf{C}_{g,t}^\mathbf{p}$ using (\ref{cg_RGPG}).

\STATE Calculate the gain matrix $\mathbf{\hat{G}}$ according to (\ref{Ghat_RGPG}). 

\STATE Set $\mathbf{y_t} = \boldsymbol{\mu}_{g,t}^\mathbf{p}$ at locations where $\mathbf{y_t}$ is missing.  Calculate mean $\boldsymbol{\mu}_{g,t}^\mathbf{f}$ by means of (\ref{mugt_RGPG}) and covariance matrix $\mathbf{C}_t^\mathbf{f}$ by means of (\ref{Cgt_RGPG}). 

 \STATE Predict for all $x_*$ which satify $x(t) < x(t_*)$ with mean $\mathbf{m}_g^*$ and variance $\mathbf{C}_g^*$ using (\ref{m*_sequential}) and (\ref{C*_sequential}) respectively.  
   
    \STATE \textbf{return}  $\mathbf{m}_g^*$, $\mathbf{C}_g^*$.
  \end{algorithmic}
  \caption{RGP-G prediction Approach}
\end{algorithm}
\subsection{Multi-task Recursive GP (RGP)}
When the network topology is unknown, the graph structure information by means of the graph filter matrix is not possible. In this case, we aim to use the recursive multi-task GP without graphs approach (RGP).  Here,  we initialize the prior function $\mathbf{f} = \mathbf{f(X})$ at time $t=0$ as,\\
    \begin{equation}
    p_0(\mathbf{\mathbf{f}}) = \mathcal{N} (\mathbf{f};\boldsymbol{\mu}_{0}^\mathbf{f}, \mathbf{C}_0^\mathbf{f})
    \end{equation}
with mean $\boldsymbol{\mu}_{0}^\mathbf{f} = \mathbf{0}$ and covariance $\mathbf{C}_0^\mathbf{f}$ defined as,  \\
\begin{equation}
\mathbf{C}_0^\mathbf{f} =(\mathbf{K}_{c} \otimes \mathbf{I}) \otimes \mathbf{K} + \sigma_{\epsilon}^2 \mathbf{I}
\label{C0_f}
\end{equation}
We define the matrices $\mathbf{A}$, $\mathbf{D}$ and $\mathbf{F}$ as,
\begin{equation}
\mathbf{A} = (\mathbf{K}_{c} \otimes \mathbf{I}) \otimes \mathbf{K} + \sigma_{\epsilon}^2 \mathbf{I},
\label{A_RGP}
\end{equation}
\begin{equation}
\mathbf{D} = (\mathbf{K}_{c} \otimes \mathbf{I}) \otimes \mathbf{K}_t + \sigma_{\epsilon}^2 \mathbf{I},
\label{D_RGP}
\end{equation}
\begin{equation}
\mathbf{F} = (\mathbf{K}_{c} \otimes \mathbf{I}) \otimes \mathcal{K}_{tt} + \sigma_{\epsilon}^2 \mathbf{I}
\label{F_RGP}
\end{equation}

Instead of the matrices (\ref{A_RGPG}), (\ref{D_RGPG}) and (\ref{F_RGPG}) defined for RGP-G approach, we use the matrices defined in (\ref{A_RGP}), (\ref{D_RGP}) and (\ref{F_RGP}).  The sequential inference and update step remains the same for both approaches.  In this approach, we can perform both the interpolation and prediction similar to the RGP-G approach. In case of RGP Interpolation approach, the predicted mean and covariance is denoted by $\mathbf{m}^*$ and $\mathbf{C}^*$ given as,
\begin{equation}
    \mathbf{m^*} = \mathbf{D}^\intercal \mathbf{A}^{-1} \boldsymbol{\mu}_{T}^\mathbf{f}
    \label{M*RGP}
\end{equation}
\begin{equation}
    \mathbf{C^*} = 
    \mathbf{B} +  \mathbf{J}_{t^*} (\mathbf{C}_{T}^\mathbf{f}) \mathbf{J}_{t^*}^\intercal
    \label{C*RGP}
\end{equation}

The RGP approach used for interpolation is summarized in Algorithm 4.
\begin{algorithm}
\KwInput{Basis vector $\mathbf{x}$, total time instants $T$, $\alpha$, sequential measurements $\mathbf{y}_t \in \mathbb{R}^{dM}$, $\mathbf{K}$, $\mathbf{K}_{c}$ }
\begin{algorithmic}[1]
 
\STATE Initialization: $\boldsymbol{\mu}_{0}^\mathbf{f} = \mathbf{0}$ and $\mathbf{C}_{0}^\mathbf{f} = (\mathbf{K}_{c} \otimes \mathbf{I}) \otimes \mathbf{K}$.
  
    \FOR{$t = 1,...,T$}
\STATE  Calculate the gain matrix $\mathbf{J}_{t}$ according to (\ref{J_RGPG}) using the matrix $\mathbf{A}$ and $\mathbf{D}$ from (\ref{A_RGP}) and (\ref{D_RGP}) respectively . 

\STATE Calculate mean $\boldsymbol{\mu}_{t}^\mathbf{p}$ using (\ref{mug_RGPG}) and covariance matrix $\mathbf{C}_{t}^\mathbf{p}$ using (\ref{cg_RGPG}).

\STATE Calculate the gain matrix $\Tilde{\mathbf{G}}_t$ according to (\ref{Ghat_RGPG}). 

\STATE Set $\mathbf{y_t} = \boldsymbol{\mu}_{t}^\mathbf{p}$ at locations where $\mathbf{y_t}$ is missing.  Calculate mean $\boldsymbol{\mu}_{t}^\mathbf{f}$ by means of (\ref{mugt_RGPG}) and covariance matrix $\mathbf{C}_{t}^\mathbf{f}$ by means of (\ref{Cgt_RGPG}). 
    \ENDFOR
    \STATE Perform imputation at time $x_*$ by means of mean $\mathbf{m}^*$ (\ref{M*RGP}) and covariance matrix $\mathbf{C}^*$ (\ref{C*RGP}) .  
    \STATE \textbf{return} $\mathbf{m}^*$, $\mathbf{C}^*$.
  \end{algorithmic}
  \caption{Multi-task RGP Approach}
\end{algorithm}
The computational complexity associated with Algorithm 2, 3 and 4  for $n$ basis vectors and $dM$ number of observations at step $t$ is $\mathcal{O}(dMn^2)$. This complexity is driven by the gain matrix calculation in (\ref{J_RGPG}). Use of recursive GP significantly reduces the computational complexity as compared to the full GP approach. 

We next show the use of graph information in RGP-G approach reduces the uncertainty of the posterior distribution when the measurements are recursively processed at time $t = 1,...,T$.
\begin{theorem}
The variance of the estimator of $\mathbf{f}$ using the RGP-G  (Algorithm 2 and 3)  of the distribution $p(\mathbf{f)}$ is less than the variance of the estimator of $\mathbf{f}$ using RGP at time $t = 0$ i.e.,
$$tr(\mathbf{C}_0^\mathbf{f}) > tr(\mathbf{C}_{g,0}^\mathbf{f})$$
where $\mathbf{C}_0^\mathbf{f}$ is defined in (\ref{C0_f}) and $\mathbf{C}_{g,0}^\mathbf{f}$ is defined in (\ref{Cg0}).  

\textit{Proof.}
In order to prove this theorem, we need to show that the trace of $\Delta \mathbf{C}_0 = \mathbf{C}_0^\mathbf{f} - \mathbf{C}_{g,0}^\mathbf{f}$ is nonnegative. The initial covariance matrix $\mathbf{C}_{g,0}^\mathbf{f}$ at time $t=0$ is defined with a graph filter $\mathbf{S}$ given in (\ref{graph_filter}). The Laplacian matrix $\mathbf{L}$ used in this graph filter has an eigen-decomposition corresponding to,
\begin{equation}
    \mathbf{L = VU_G V^\intercal}
\end{equation}
where, $\mathbf{U_G} = diag(U(1), U(2),.., U(M))$ and $\mathbf{V}$ denote the diagonal eigenvalue matrix and the associated eigenvectors respectively. Every eigenvalue of the laplacian matrix $\mathbf{L}$ is non-negative \cite{newman2018networks}. 
We need to prove,
\begin{equation}
    tr(\mathbf{C}_0^\mathbf{f} - \mathbf{C}_{g,0}^\mathbf{f}) \geq 0
\end{equation}
$$tr((\mathbf{K}_{c} \otimes \mathbf{I}) \otimes \mathbf{K} - (\mathbf{K}_{c} \otimes \mathbf{S}^2) \otimes \mathbf{K} ) \geq 0, $$
The trace of the Kronecker product of three matrices is the product of the traces of the matrices. Hence, we get,
$$ tr(\mathbf{K}_{c}) tr(\mathbf{I}) tr(\mathbf{K}) - (tr(\mathbf{K}_{c}) tr(\mathbf{S}^2) tr(\mathbf{K})) \geq 0, $$
The kernel matrix $\mathbf{K}$ and $\mathbf{K}_{c}$ is positive semidefinite by construction. Hence, we need to prove,
$$tr(\mathbf{I} - \mathbf{S^2}) > 0,$$
$$tr(\mathbf{I} - \mathbf{V (I + \alpha U_G)^{-2}V^\intercal) \geq 0 },$$
$$\sum_{i=1}^M \Bigg( 1 - \frac{1}{(1 + \alpha U(i))^2}\Bigg) \geq 0.$$
Let $U(s)$ denote the smallest non-zero eigenvalue of $\mathbf{L}$, we then need to prove
\begin{equation*}
(1 + \alpha U(s))^{-2} \leq 1,
\end{equation*}
\begin{equation*}
(1 + \alpha U(s)) \geq 1
\end{equation*}
\begin{equation*}
\alpha U(s) \geq 0.
\end{equation*}
For $\alpha > 0$, the smallest non-zero eigenvalue of $\mathbf{L} \enspace \text{i.e.}, U(s) > 0$.
Hence, $tr(\mathbf{C}_0^\mathbf{f} - \mathbf{C}_{g,0}^\mathbf{f}) > 0.$ $\blacksquare$
\end{theorem}

\begin{theorem}
The posterior covariance matrix of the estimator of function $\mathbf{f}$ for RGP-G approach at time $t$ i.e.,  $\mathbf{C}^*_g$ in (\ref{C*RGPG}) is smaller than the posterior covariance matrix for the RGP based estimator of $\mathbf{f}$ given as $\mathbf{C^*}$ in (\ref{C*RGP}) evaluated without graph information $\forall{t =1,...,T}.$

\textit{Proof:} We need to prove,
$tr(\mathbf{C}_g^*) < tr(\mathbf{C}^*)$ for all $t = 1,...,T$, where
\begin{equation*}
  \mathbf{C}^* = \mathbf{B} +  \mathbf{J}_{t*} (\mathbf{C}^\mathbf{f}_T) \mathbf{J}_{t*}^\intercal  
\end{equation*}
and $\mathbf{C}_g^*$ is defined in (\ref{C*RGPG}).
From Theorem 1, we have proved that $\Delta \mathbf{C}_0 = tr(\mathbf{C}_0^\mathbf{f} - \mathbf{C}_{g,0}^\mathbf{f}) > 0$.

The matrix $\mathbf{B}$ for RGP approach can be defined as,
\begin{equation*}
\mathbf{B} =\mathbf{F - D^\intercal A^{-1} D}
\end{equation*}
Similarly, we have $\mathbf{B}_g$ for RGP-G approach as defined in (\ref{B_g_RGPG}). 
Thus, we have, 
\begin{equation}
\begin{aligned}
tr(\mathbf{B - B}_g) = tr((\mathbf{F- F_g}) -   (\mathbf{D - D}_g)^\intercal (\mathbf{C}_0^\mathbf{f} - \mathbf{C}_{g,0}^\mathbf{f})^{-1} \\
(\mathbf{D - D}_g))
\end{aligned}
\end{equation}
As $tr(\mathbf{C}_0^\mathbf{f} - \mathbf{C}_{g,0}^\mathbf{f}) > 0$ and the Schur complement of a positive definite matrix is also positive definite. Therefore, we have 
\begin{equation*}
    tr(\mathbf{B - B}_g) > 0
\end{equation*}
Similarly,
\begin{equation}
\begin{aligned}
    tr(\mathbf{C}_t^p - \mathbf{C}_{g,t}^{\mathbf{p}}) = tr\big((\mathbf{B - B}_g) + (\mathbf{J}_t- \mathbf{J}_{g,t})(\mathbf{C}_{g,t-1}^\mathbf{f} - \mathbf{C}_{t-1}^\mathbf{f}) \\
    \times (\mathbf{J}_t- \mathbf{J}_{g,t})^\intercal \big)
\end{aligned}
\end{equation}

At $t =1$ we have $\mathbf{C}_{0}^\mathbf{f} - \mathbf{C}_{g,0}^\mathbf{f} > 0$ (proved in theorem 1) and $tr(\mathbf{B - B}_g) > 0$. Hence,
\begin{equation*}
     tr(\mathbf{C}_1^{\mathbf{p}} - \mathbf{C}_{g,1}^{\mathbf{p}}) > 0
\end{equation*}
Similarly,  $tr(\mathbf{C}_t^{\mathbf{f}} - \mathbf{C}_{g,t}^{\mathbf{f}}) > 0 \enspace \forall{t=1,...,T}$. The posterior distribution of the imputations 
\begin{equation*}
    tr(\mathbf{C^*} - \mathbf{C^*_g}) > 0. \enspace \blacksquare
\end{equation*}
\end{theorem}

Next, we discuss the robustness of the graph filter to topology uncertainties.
\subsection{Robustness to topology uncertainties}
The distribution grid topology can be unreliable and incorrectly estimated \cite{gandluru2019joint}, \cite{xu2021adaptive}. Thus, it would be desirable for our GP predictions to be robust against the uncertainties in topology. The following discussion presents the conditions under which the proposed graph filter will be stable to perturbations in the graph topology, as proved in \cite{kenlay2021interpretable}. The graph filter is said to be stable against a perturbation if the perturbation does not lead to large changes in the filter output. 
The graph Laplacian matrix $\mathbf{L}$ is also known as the graph shift operator (GSO) which can be decomposed as $\mathbf{L = V U_G V^\intercal}$, where $\mathbf{U_G}$ and $\mathbf{V}$  denote the eigenvalues and the eigenvectors of the Laplacian matrix, respectively. Let the Laplacian matrix of the perturbed graph $\mathcal{G}_p$ be $\mathbf{L}_p$. The magnitude of the error matrix is defined as  $\|\mathbf{E} \|_2 = \|\mathbf{L} - \mathbf{L}_p\|_2$.
In \cite{kenlay2021interpretable}, a graph filter $g$ is said to be linearly stable for any GSO if, for any GSO  $\Delta$ and  $\Delta_p$, we have the following conditions satisfied, i.e.,
\begin{equation}
  \|g(\mathbf{L}) - g(\mathbf{L}_p) \|_2 \leq C \|\mathbf{E} \|_2 
  \label{bounds}
\end{equation}
where, $C$ is a positive constant. The graph filter defined in (\ref{graph_filter}) can also be written as, 
 \begin{equation}
   \mathbf{y_d}^* = \mathbf{(I}_M + \alpha \mathbf{L})^{-1} \mathbf{y} = \sum_{l=0}^{M-1} \frac{1}{1+ \alpha U_l} \langle y ,\mathbf{V}_l\rangle \mathbf{V}_l
\end{equation}
where, $\langle y,\mathbf{V}_l\rangle$ is the graph Fourier transform of $y$ on the vertices of the graph. Equivalently, $\mathbf{y_d}^* = \hat{h} (L) \mathbf{y}$, where $\hat{h} (L)  = \frac{1}{1 + \alpha U}$ can be viewed as low-pass filter.
The low-pass filter is said to be linearly stable for perturbations in the graph as it satisfies (\ref{bounds}). This property was proven in \cite{kenlay2021interpretable}  and given as, 
\begin{equation}
  g(\mathbf{L}) - g(\mathbf{L}_p) \|_2 \leq \alpha \|\mathbf{E} \|_2. 
  \label{eq1}
\end{equation}
Complete analysis of the impact of topology uncertainties on graph filter stability and eventually the RGP-G approach performance will be pursued as part of our future work.
\section{Matrix completion based DSSE}
 While not the primary focus or contribution of our work, we  provide a brief summary of the matrix completion based DSSE proposed in \cite{donti2019matrix} for the sake of completeness. Unlike \cite{donti2019matrix}, \cite{9247106} that assumes time synchronized subset of measurements, here the GP based reconciled measurements are used within the Matrix completion (MC) based DSSE.
 MC based DSSE estimate the spatial states of the network (i.e., the voltage phasors and power injections of all the buses at a single instant of time) by exploiting the sparsity of raw measurements. Specifically, matrix completion aims to estimate the complete matrix $\mathbf{X} $ from an incomplete and noisy observation matrix by suitable low rank approximation. 

The consistent multi time-scale measurements are limited to specific spatial locations in the network where measurements are aggregated.
Assume that the measurements at the slack bus are known. Thus, we use the measurements at the non-slack buses to construct a data matrix.  Let $m$ denote the set of phases at all the non-slack buses. The noisy matrix \textbf{Z} is constructed such that each row represents a phase and each column represents the measurement associated with the phase of each bus. For each $b \in m$, each row of the matrix $\mathbf{Z} \in \mathbb{R}^{m \times n}$, $n=5$ is structured as,
\begin{equation}
[\mathbf{P}_b,\mathbf{Q}_b , \Re(\mathbf{v}_b),\Im(\mathbf{v}_b ),|\mathbf{v}_b | ], 
       \label{matrixeqn}
\end{equation}
where, $\mathbf{P}_b$ and $\mathbf{Q}_b$ represent the active power and reactive power injections at each phase of non-slack bus $b$ respectively. The terms $\Re({\mathbf{v}_b})$ and $\Im({\mathbf{v}_b})$ represent the real and imaginary parts of voltage phasors at each phase of non-slack buses respectively.
Let $\Omega \subseteq \{1,...,m\} \times \{1,...,n \}$ describe the known entries in $\mathbf{Z}$. The observation matrix $P_{\Omega}(\mathbf{Z})$ is represented as, 
\begin{align}
  [P_{\Omega}(\textbf{Z})]_{mn} =
  \begin{cases}
  \mathbf{Z}_{mn} , \enspace \text{if} \enspace (m,n) \in \Omega 
  \\
  0, \enspace \enspace\text{otherwise}
  \end{cases}
  \label{observation_eq}
\end{align}

The matrix completion formulation (\ref{equation9}) recovers the complete low-rank matrix, as
\begin{equation}
\begin{aligned}
\mathbf{\hat{X} }= 
\underset{\mathbf{X} }{\text{\enspace argmin}}
\enspace \| \mathbf{X} \|_* \\
\text{subject to} 
&  & \|P_{\Omega}(\mathbf{Z}) - P_{\Omega}(\mathbf{X}) \|_F^2 < \epsilon\\ 
\end{aligned}
\label{equation9}
\end{equation}
\begin{equation}
    \mathbf{v} = \mathbf{M} \begin{bmatrix}\mathbf{P}\\ \mathbf{Q} \end{bmatrix} + \mathbf{w},
    \label{equ_pf}
\end{equation}
\begin{equation}
\mathbf{|v| = K\mathbf} \begin{bmatrix}\mathbf{P}\\ \mathbf{Q} \end{bmatrix} + \mathbf{|w|},
\label{eq4}
\end{equation}
Here, the nuclear norm $\| \mathbf{X} \|_*$ is the sum of the singular values of the matrix $\mathbf{X}$.
 (\ref{equ_pf}) and (\ref{eq4}) captures the linearized power-flow constraint relating voltage phasors ${\mathbf{v}}$ and voltage magnitude ${\mathbf{|v|}}$ to the power measurements as given in \cite{bernstein2017linear}. 
More details about the matrix completion based DSSE can be found in \cite{donti2019matrix}, \cite{9247106}. Additionally, the impact of uncertain topology on matrix completion based DSSE is considered in \cite{karimi2021joint}, \cite{9247106}. Additionally, a more comprehensive integrated robustness analysis of graph filter and matrix completion based DSSE will be considered in our future work.

\section{Simulation Results}
The efficacy of the proposed approach is verified on the three-phase unbalanced IEEE 37 bus \cite{malekpour2015radial}, and IEEE 123 bus test system \cite{schneider2017analytic}. An aggregated 24-hr load profile at the primary nodes consists of a mixture of load profiles, i.e., industrial and commercial load profiles obtained from \cite{carmona2013fast}, and residential loads obtained from \cite{al2016state}. Reactive power profiles are obtained by assuming a power factor of 0.9 lagging.
Other profiles at different nodes were obtained by adding a random noise term and a sinusoidal wave of random amplitude spanning the 24-hr period.
By utilizing this data, the voltage profile at all nodes is obtained by running load flow. The aggregated smart meter data are averaged over 15-min intervals while the voltage magnitude measurements are sampled at a 1-min interval. Thus, we have considered two sensor types for the case study. We have assumed RBF kernel for all the GP-based approaches. The imputation is performed for the aggregated smart meter data at a 1-min interval. We compare the performance of the proposed RGP-G Interpolation against the linear interpolation approach \cite{gomez2014state}, RGP (Algorithm 3), and full GP (Algorithm 1) approach.  The RGP-G prediction approach is compared with \cite{karimipour2015extended}.  Algorithms 2, 3, and 4 are initialized using their respective mean and co-variance function associated with the GP function at time $t =0$. Here, the hyper-parameters associated with the GP function can be obtained by either training the proposed approaches using historical data or using cross-validation techniques. The hyper-parameters involved in the proposed approach are $\theta = [l, \sigma_s^2, \sigma_{\epsilon}^2]$, where $l,\sigma_s^2, \sigma_{\epsilon}^2, $ are defined in (\ref{hyperparameter}). We have used the grid search method guided by a five-fold cross-validation technique to obtain the hyper-parameters for our problem. In the cross-fold validation technique, one fold of the measurement set is retained as a validation set, and the other folds as a training set. Each time a different set is chosen as the validation set, and this procedure is repeated five times.   We select a finite set of reasonable hyper-parameter values to perform a grid search. The performance of each combination is evaluated through cross-validation on the training set. This approach evaluates the MAPE for each possible combination of hyperparameter values and chooses the set that minimizes the error on the validation set.   More details on the grid-search-based cross-fold validation technique for Gaussian process hyperparameter tuning can be found in \cite{rasmussen2003gaussian}. Another approach is to consider the historical data for hyperparameter tuning. The historical measurements of multi time-scale measurements can be used to obtain the hyper-parameters by maximizing the log marginal likelihood of the historical time-series data. The log-likelihood can be computed in closed form as given in \cite{rasmussen2003gaussian}. It is important to note that the proposed approach does not require any extra training set for imputation.  The parameter $\alpha$ for the RGP-G  approach is set to 0.05. There are three cases by which we illustrate the performance of the multi-task RGP-G approach.
\begin{enumerate}
\item \underline{Case 1:} In this case, we consider the measurement noise as mean 0 with standard deviation equal to 1\% of the actual values. Fig. \ref{fig:rgpg_0pcresult_node3_ver1}  shows the performance of the RGP-G interpolation approach at 0\% missing measurement case for an IEEE 37 bus test system. Here, the time-series is the active power injection at node 11 of phase A. The RGP-G interpolation approach recursively updates the GP function in the 24-hr batch and later performs imputation at 1-min interval. The 95\% confidence interval indicates the uncertainty bounds associated with the imputed measurements. The ideal case, i.e., 0\% missing measurements dataset, has no missing measurements, but
the dataset consists of a subset of the total measurements,
representing an unobservable condition. For instance, let
us assume that the aggregated AMI measurements in this
dataset are available at 15-min intervals. Therefore, if we
consider a 24-hr duration to perform imputations every
minute, we have only 96 measurements per AMI sensor
out of the total 1440-time instances.

\begin{figure}[h!]
\centering
\includegraphics[width= 0.55\textwidth]{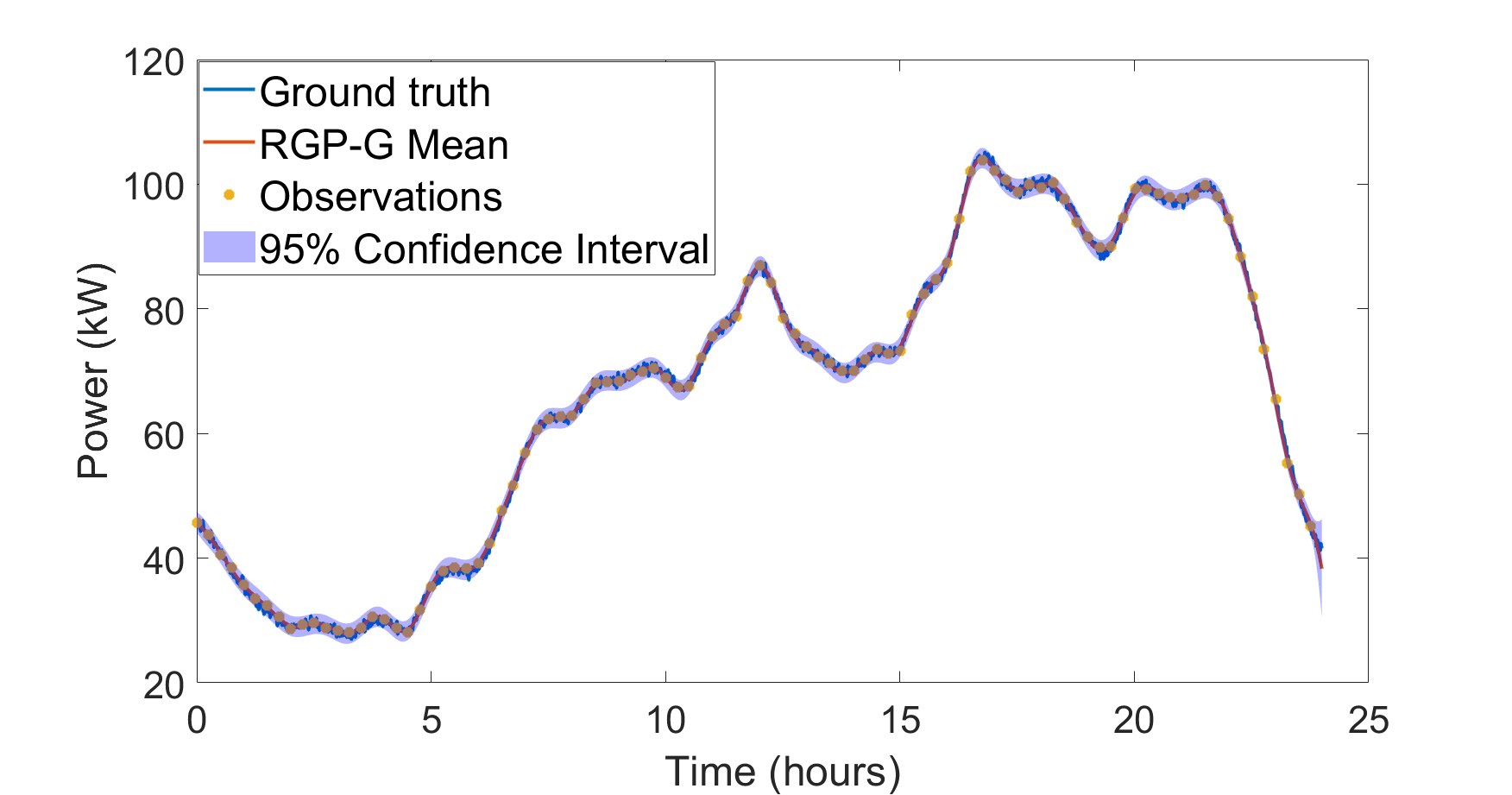}
   \caption{{RGP-G Interpolation approach of active power injection timeseries at node 11, Phase A}}
     \label{fig:rgpg_0pcresult_node3_ver1}
 \end{figure}

\item \underline{Case 2:} In this case, we perform the reconciliation and state estimation using the noisy time-series measurements corrupted by 0 mean and 10\% standard deviation  for a total of four hours duration. In this case, we fix the number of sensors and their locations corresponding to a particular FAD. The meters are placed randomly in the network. Fig.\ref{fig:meter_deployment} shows the meter deployment for IEEE 37 bus test system for 50\% FAD. Thus, there are no time-series measurements at the nodes where the sensors are absent.  We then perform imputation using these incomplete measurements.
Fig.\ref{compare_noisycase} shows the comparison of the RGP-G approach with linear interpolation. It can be seen that the former approach provides smoother imputation than the latter one. We compare the performance of all the five approaches using the mean absolute percentage error (MAPE) metric. Tables \ref{TableIEEE37} and \ref{TableIEEE123} tabulate their performances for IEEE 37 and IEEE 123 bus test systems, respectively.  It can be inferred that the performance of the RGP-G interpolation is superior to the other four approaches. In the linear interpolation approach, each time-series data is imputed individually without exploiting any spatio-temporal property of the data. In contrast, the full GP approach (Algorithm 1) exploits temporal relationships for imputation. The GP function update is performed using all the measurements in batch mode at once, which is computationally expensive.

The proposed approach can impute at any missing measurement level and the uncertainty as well as MAPE will increase with the increase in percentage of missing measurements. Uncertain imputed measurements affects the state estimation process. The knowledge of the uncertainty in imputations is used to guide the DSSE process using a Bayesian MC framework as proposed in \cite{dahale2022bayesian}. Analytical bounds on the estimation error of matrix completion approach in the presence of missing and noisy measurements are derived in \cite{5948421}. Since the primary goal of our work is to introduce the novel RGP-G and RGP algorithms as viable options to deal with multi time-scale measurements, derivations of error bounds based on \cite{5948421} will be pursued as a part of the future work.

\begin{figure}[h!]
\centering
\includegraphics[width= 0.25\textwidth]{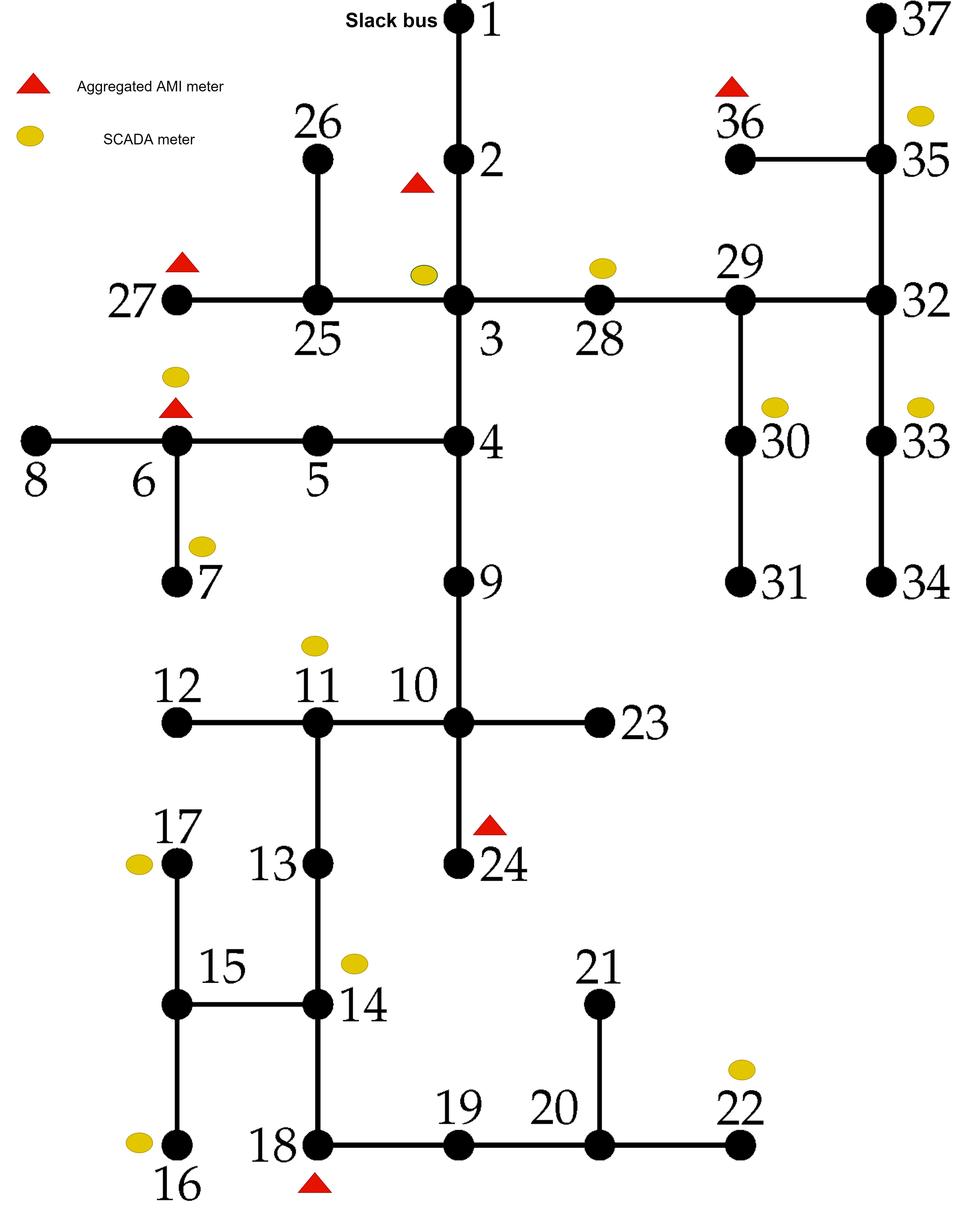}
   \caption{{Meter deployment corresponding to 50\% FAD for IEEE 37 bus test system}}
     \label{fig:meter_deployment}
 \end{figure}
 
\begin{figure}[h!]
\centering
\includegraphics[width= 0.55\textwidth]{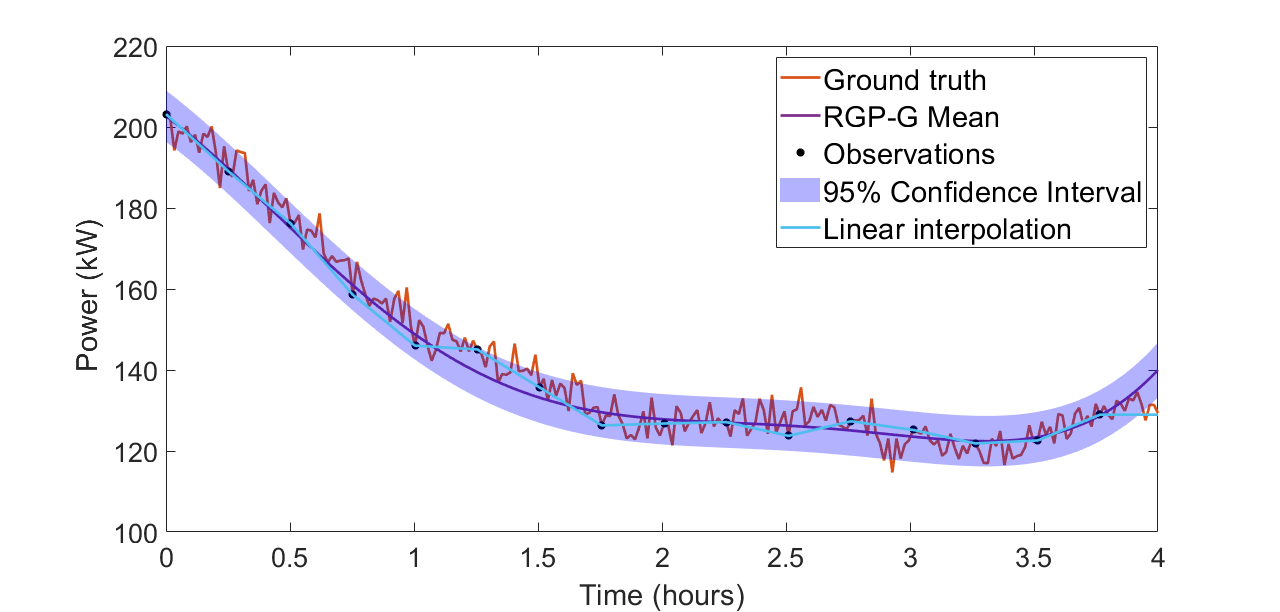}
  \caption{{Comparison of RGP-G interpolation and linear interpolation approach of an active power injection time-series at Node 2 of Phase A}}
    \label{compare_noisycase}
 \end{figure}

\begin{table}
\centering
\caption{Case 2: MAPE of active and reactive power imputed time-series data (IEEE 37 bus test system)}
\begin{tabular}{|l|l|l|l|l|} 
\hline
Scenario                                                    & \begin{tabular}[c]{@{}l@{}}Proposed\\RGP-G I\\\end{tabular}        & RGP I                                                          & \begin{tabular}[c]{@{}l@{}}Full GP\\\end{tabular}              & Linear I                                                           \\ 
\hline
\begin{tabular}[c]{@{}l@{}}0\% \\ missing\end{tabular}  & \begin{tabular}[c]{@{}l@{}}1.4\% (P)\\2.3\% (Q)\end{tabular}       & \begin{tabular}[c]{@{}l@{}}1.8\%~(P) \\2.6\%~ (Q)\end{tabular} & \begin{tabular}[c]{@{}l@{}}8.5\%~ (P)\\5.3\%~(Q)\end{tabular}  & \begin{tabular}[c]{@{}l@{}}3.18\%~~(P)\\6.5\%~ ~~(Q)\end{tabular}  \\ 
\hline
\begin{tabular}[c]{@{}l@{}}10\%\\ missing\end{tabular}  & \begin{tabular}[c]{@{}l@{}}3.23\%~ (P)\\3.4\%~ ~ (Q)\end{tabular}  & \begin{tabular}[c]{@{}l@{}}4.2\%~(P) \\5.35\%~(Q)\end{tabular} & \begin{tabular}[c]{@{}l@{}}8.1\%~(P) \\8.7\%~(Q)\end{tabular}  & \begin{tabular}[c]{@{}l@{}}3.37\%~(P)\\7.72\%~~(Q)\end{tabular}    \\ 
\hline
\begin{tabular}[c]{@{}l@{}}20\% \\ missing\end{tabular} & \begin{tabular}[c]{@{}l@{}}3.02\%~~(P)\\3.07\%~ ~(Q)~\end{tabular} & \begin{tabular}[c]{@{}l@{}}4.2\%~(P) \\9.89\%~(Q)\end{tabular} & \begin{tabular}[c]{@{}l@{}}7.8\%~(P) \\8.9\%~~(Q)\end{tabular} & \begin{tabular}[c]{@{}l@{}}3.6\%~(P) \\8.1\%~~(Q)\end{tabular}     \\
\hline
\end{tabular}
\label{TableIEEE37}
\end{table}
\begin{table}
\centering
\caption{Case 2: MAPE of active and reactive power imputed time-series data (IEEE 123 bus test system)}
\begin{tabular}{|l|l|l|l|l|} 
\hline
Scenario                                                    & \begin{tabular}[c]{@{}l@{}}Proposed\\RGP-G I\\\end{tabular}    & RGP I                                                           & \begin{tabular}[c]{@{}l@{}}Full \\ GP\\\end{tabular}              & Linear I                                                          \\ 
\hline
\begin{tabular}[c]{@{}l@{}}0\% \\ missing\end{tabular}  & \begin{tabular}[c]{@{}l@{}}4.11\% (P)\\1.7\%~ (Q)\end{tabular} & \begin{tabular}[c]{@{}l@{}}4.38\%~(P)\\1.72\% (Q)\end{tabular}  & \begin{tabular}[c]{@{}l@{}}8.9\%~~(P)\\5.68\%~(Q)\end{tabular}    & \begin{tabular}[c]{@{}l@{}}6.67\%~(P)\\2.5\%~(Q)\end{tabular}     \\ 
\hline
\begin{tabular}[c]{@{}l@{}}10\%\\ missing\end{tabular}  & \begin{tabular}[c]{@{}l@{}}4.45\% (P)\\1.8\%~~(Q)\end{tabular} & \begin{tabular}[c]{@{}l@{}}5.1\%~(P)\\2.5\% (Q)\end{tabular}    & \begin{tabular}[c]{@{}l@{}}15.6\%~(P)\\2.7\%~ ~(Q)\end{tabular}   & \begin{tabular}[c]{@{}l@{}}6.75\%~(P)\\10.3\%~(Q)\end{tabular}    \\ 
\hline
\begin{tabular}[c]{@{}l@{}}20\% \\ missing\end{tabular} & \begin{tabular}[c]{@{}l@{}}4.6\%~~(P)\\2.18\%~(Q)\end{tabular} & \begin{tabular}[c]{@{}l@{}}5.41\%~(P)\\2.9\%~ ~(Q)\end{tabular} & \begin{tabular}[c]{@{}l@{}}15.8\%~~(P)\\10.48\%~ (Q)\end{tabular} & \begin{tabular}[c]{@{}l@{}}17.87\%~(P)\\12.28\%~(Q)\end{tabular}  \\
\hline
\end{tabular}
\label{TableIEEE123}
\end{table}
The proposed recursive GP approaches assumes that the measurement data is corrupted by Gaussian noise as seen from (\ref{ytd_rgpg}). Also, the optimization formulation in (\ref{eq14}) assumes Gaussian noise. These are common assumptions used in many prior efforts \cite{venkitaraman2020gaussian, shuman2013emerging, rasmussen2003gaussian}. The proposed approach in its current form can be applied to non-Gaussian noise but will not be optimal. 
Table \ref{TableLaplacian} shows the performance of the RGP-G approach with Laplacian noise with 0 mean with standard deviation equal to 5\%, 10\% of the actual power values. As seen from Table \ref{TableLaplacian}, it can be inferred that the performance of the recursive GP approach under non-Gaussian noise scenarios is not optimal.
Alternately, under a non-Gaussian measurement noise scenario, a warped GP approach \cite{snelson2003warped} can be used. In this approach, the observations are transformed into a latent space such that the transformed data has Gaussian noise and will be better modeled by the GP. Developing a multi-task recursive GP framework using the warped GP will be pursued as part of our future work.

\begin{table}[h!]
\caption{MAPE of proposed RGP-G approach for Laplacian and Gaussian noise with standard deviation set as percentage of actual power values}  
\centering
\begin{tabular}{|l|l|l|}
\hline
Standard deviation of noise & 5\%    & 10\%    \\ \hline
Laplacian  noise            & 4.14\% & 12.47\% \\ \hline
Gaussian  noise             & 2.6\%  & 3.07\%  \\ \hline
\end{tabular}
\label{TableLaplacian}
\end{table}
The consistent time-series measurements are further used to estimate the states using the matrix completion-based DSSE approach discussed in section IV. While performing the matrix completion for a particular FAD, the corresponding entries in the measurements matrix are zero,  with no sensor measurements.  Table \ref{TableMC} shows the absolute errors and relative error reductions for RGP-G and linear interpolation methods. It can be deduced that RGP-G based technique significantly reduces error at all FADs. For example, the error in estimating reactive power using GP-based imputed time-series is reduced by 46\% at 90\% FAD compared to the linearly interpolated time-series. The reduction in voltage state estimation error is more modest mainly due to the robustness of matrix completion based DSSE that includes the topology information. 

\item \underline{Case 3:} In this case, we compare the RGP-G prediction aided matrix completion with  \cite{karimipour2015extended}.  The method in \cite{karimipour2015extended} uses a data collation method to reconcile heterogeneous measurements and a Kalman filter method to perform DSSE. The data collation consists of an exponential moving average method to extrapolate the slow-rate measurements. 
Table \ref{TableRGPGPrediction} shows the prediction errors for different percentages of missing temporal data.
Our proposed approach has several advantages over \cite{karimipour2015extended}. Firstly, the DSSE in \cite{karimipour2015extended}  requires the measurement redundancy  (ratio of number of measurements to the total states) higher than two. Thus, unlike our proposed approach, the method in \cite{karimipour2015extended} fails in low-observable conditions.  Secondly, it can be observed that the imputation error in \cite{karimipour2015extended} is higher than the proposed RGP-G prediction approach as seen from Table \ref{TableRGPGPrediction}.

\end{enumerate}

\begin{table*}[t]
\centering
\caption{Absolute errors and relative error reductions (\%) compared to the actual measurements}
\begin{tabular}{|l|l|l|l|l|l|l|l|l|l|}
\hline
\multicolumn{1}{|c|}{Scenario}                                                          & \multicolumn{3}{c|}{FAD = 50\%}                                                                                                         & \multicolumn{3}{c|}{FAD = 70\%}                                                                                                         & \multicolumn{3}{c|}{FAD = 90\%}                                                                                                         \\ \hline
\multicolumn{1}{|c|}{\begin{tabular}[c]{@{}c@{}}Estimated \\ measurements\end{tabular}} & \multicolumn{1}{c|}{Linear} & \multicolumn{1}{c|}{RGP-G} & \multicolumn{1}{c|}{\begin{tabular}[c]{@{}c@{}}\%\\ reductions\end{tabular}} & \multicolumn{1}{c|}{Linear} & \multicolumn{1}{c|}{RGP-G} & \multicolumn{1}{c|}{\begin{tabular}[c]{@{}c@{}}\%\\ reductions\end{tabular}} & \multicolumn{1}{c|}{Linear} & \multicolumn{1}{c|}{RGP-G} & \multicolumn{1}{c|}{\begin{tabular}[c]{@{}c@{}}\%\\ reductions\end{tabular}} \\ \hline
\begin{tabular}[c]{@{}l@{}}Active \\ power (kW)\end{tabular}                            & 8.3                         & 7.9                        & 5.1\%                                                                         & 8.35                        & 7.5                        & 11.33\%                                                                         & 2.8                         & 2.4                        & 16.67\%                                                                         \\ \hline
\begin{tabular}[c]{@{}l@{}}Reactive \\ power (kVAR)\end{tabular}                        & 3.87                        & 3.46                       & 11.8\%                                                                          & 3.8                         & 3.1                        & 22.5\%                                                                          & 1.57                        & 1.07                       & 46.7\%                                                                         \\ \hline
\begin{tabular}[c]{@{}l@{}}Voltage \\ magnitude (p.u)\end{tabular}                      & 0.85                        & 0.84                       & 1.19\%                                                                        & 0.85                        & 0.83                       & 2.41\%                                                                          & 0.22                        & 0.18                       & 22.22\%                                                                         \\ \hline
\end{tabular}
\label{TableMC}
\end{table*}

\begin{table}
\centering
\caption{Case 3: MAPE of imputed time-series data of active and reactive power (IEEE 37 bus test system)}
\begin{tabular}{|l|l|l|} 
\hline
Scenario                                                    & \begin{tabular}[c]{@{}l@{}}Proposed\\RGP-G\\Prediction\\\end{tabular} & \begin{tabular}[c]{@{}l@{}}Data Collation\\method \cite{karimipour2015extended}\end{tabular}  \\ 
\hline
\begin{tabular}[c]{@{}l@{}}0\% \\ missing\end{tabular}  & \begin{tabular}[c]{@{}l@{}}2.26\% (P)\\2.87\% (Q)\end{tabular}                         & \begin{tabular}[c]{@{}l@{}}6.44\%~ (P)\\6.75\%~~(Q)\end{tabular}                      \\ 
\hline
\begin{tabular}[c]{@{}l@{}}10\%\\ missing\end{tabular}  & \begin{tabular}[c]{@{}l@{}}3.5\%~ (P)\\5.65\%~(Q)\end{tabular}                         & \begin{tabular}[c]{@{}l@{}}7.75\%~ (P)\\7.16\%~~(Q)\end{tabular}                      \\ 
\hline
\begin{tabular}[c]{@{}l@{}}20\% \\ missing\end{tabular} & \begin{tabular}[c]{@{}l@{}}4.8\%~(P)\\7.3\% (Q)\end{tabular}                           & \begin{tabular}[c]{@{}l@{}}8.13\%~ (P)\\8.86\%~~(Q)\end{tabular}                      \\
\hline
\end{tabular}
\label{TableRGPGPrediction}
\end{table}

We now discuss the  scalability of the proposed approach for large test systems. 
\subsection{Scalability Analysis}
As discussed in section III, the computational complexity associated with Algorithm 2, 3 and 4 for $M$ node system with $d$ different sensor data streams and $n$ time instants will be of the order of $\mathcal{O}(dMn^2)$. Hence, while the approach can be used for $M$ = 8500, the complexity grow as $M$ increases. To address the scalability issue, a distributed implementation of the proposed multi-task recursive GP approach is possible. 
We perform the distributed implementation on the 11,000-node feeder proposed in \cite{8879622}. The  11,000-node test
feeder is constructed by connecting an IEEE 8,500-node test feeder and an EPRI Ckt7 test feeder at the substation. To perform the distributed recursive GP approach, we partition the 11,000 node network into four areas. We assume that smart meter measurements are available at 30\% of nodes in each area. The load profiles assigned to each node consist of industrial, residential, and commercial load profiles and and scaled according to their base loads provided in \cite{8879622}. Reactive power profiles are obtained by assuming a power factor randomly varying between 0.9 and 0.95 lagging. The smart meter measurements are averaged over 15-minute intervals with measurement noise as mean 0 and standard deviation equal to 1\% of the actual power values.  We perform the multi-task RGP Interpolation approach in each area for 4 hours. The MAPE for each area is tabulated in Table \ref{Table11000}. 

\begin{table}[h!]
\centering
\caption{MAPE of active and reactive power imputed time-series data (11,000 node feeder)}
\begin{tabular}{|l|l|}
\hline
Area   & MAPE  (\%) \\ \hline
 Area 1 &  1.66\%           \\ \hline
 Area 2 &  1.5\%            \\ \hline
Area 3 &  1.59\%            \\ \hline
 Area 4 &  2.26\%            \\ \hline
\end{tabular}
\label{Table11000}
\end{table}

\section{Conclusion and Future Work}
This paper proposes a recursive Gaussian process with graphs for effectively aggregating heterogeneous intermittent time-series data and using it to estimate the distribution system states in low observability conditions. The proposed approach leverages the graphical structure of the network for accurately imputing the multi time-scale measurements. It has the flexibility to perform imputations in batch mode or real-time mode. Superior imputation performance of the active and reactive power time-series measurements are obtained with the proposed approach. Further, state estimation in  IEEE  37 and the IEEE  123  bus system reveals that the power and voltage states are recovered with high fidelity.

Our proposed novel approach has significant strengths and certain limitations which will be addressed as part of our future efforts. These limitations include:
\begin{enumerate}
    \item The proposed approach is sensitive to outliers in the measurement data. Hence, we aim to develop a robust Gaussian process framework against outliers as a part of our future work. 
 \item The performance of the proposed approach depends on the hyper-parameter values of the GP function. Hence, future work will involve recursively learning the hyper-parameters as the GP functions are updated. 
 \item The proposed GP function assumes that measurement data is distributed as multivariate Gaussian. Developing a multi-task recursive GP approach for non-Gaussian noise will be pursued as a part of our future work. 
\end{enumerate}
\bibliographystyle{IEEEtran} 
\bibliography{ref.bib}

\begin{IEEEbiography}[{\includegraphics[width=1in,height=1.25in,clip,keepaspectratio]{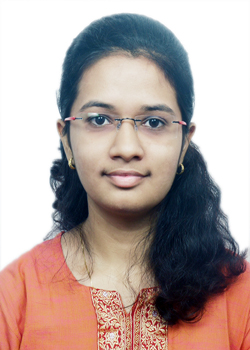}}]{Shweta Dahale} (S’20) received the B.Tech. degree in Electrical Engineering from College of Engineering, Pune, India, in
2016, and the M. Tech. degree in electrical engineering from Indian Institute of Technology
Gandhinagar, Gujarat, India, in 2018. She is currently a Ph.D. candidate at the Kansas State University, Manhattan, KS, USA.
Her research interests include optimization, machine learning and state estimation in smart grids.
\end{IEEEbiography}

\begin{IEEEbiography}[{\includegraphics[width=1in,height=1.25in,clip,keepaspectratio]{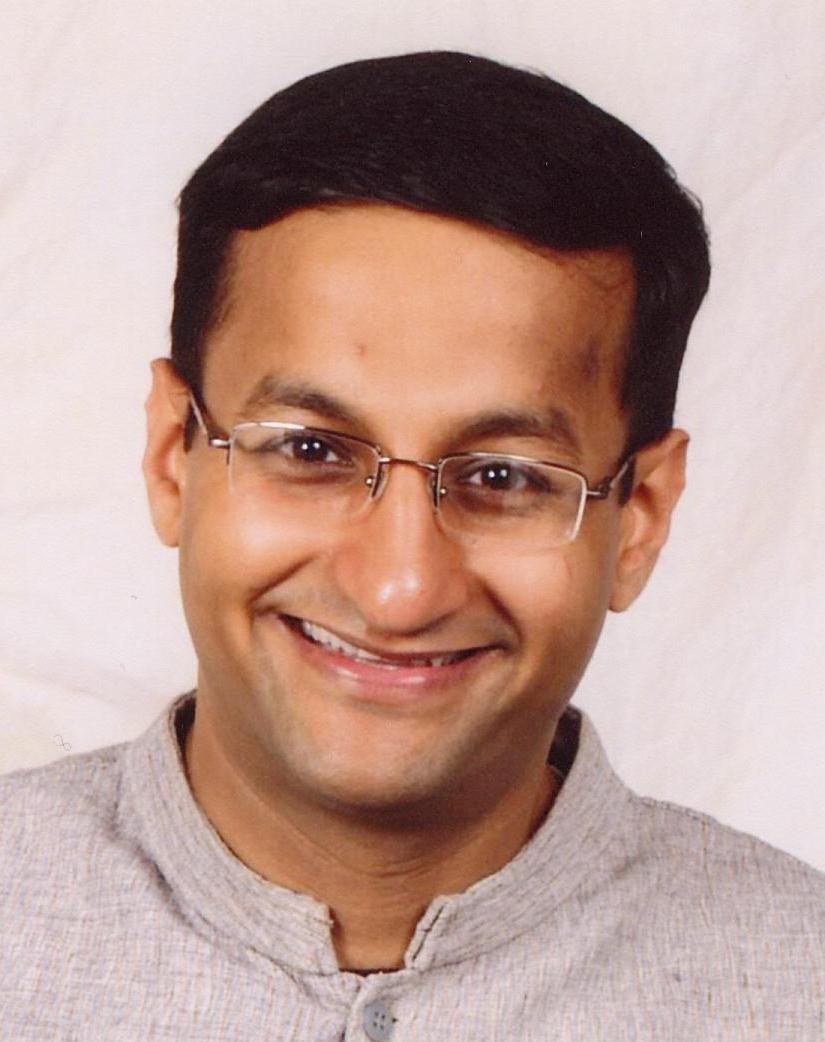}}]{Balasubramaniam
Natarajan} (SM’08) received
the B.E. degree (Hons.) in electrical and electronics engineering from Birla Institute of Technology and Science, Pilani, India, Ph.D. degree in electrical engineering from Colorado State University, Fort Collins, CO, USA, Ph.D. degree in Statistics from Kansas State University, Manhattan, KS, USA,  in 1997, 2002, and 2018, respectively.
He is currently a Clair N. Palmer and Sara M. Palmer Endowed Professor and the Director of the Wireless Communication and Information Processing Research Group. His research interests include statistical signal processing, stochastic modeling, optimization, and control theories. He has worked on and published extensively on modeling, analysis and networked estimation and control of smart distribution grids and cyber physical systems in general.  He has published over 200 refereed journal and conference articles and has served on the editorial board of multiple IEEE journals including IEEE Transactions on Wireless Communications. 
\end{IEEEbiography}

\end{document}